\documentclass[aps,prb,groupedaddress,twocolumn,showpacs]{revtex4-1}

\usepackage{relsize}
\usepackage{amsmath,amssymb,bbm}
\usepackage{graphicx}

\newcommand{\dm}{d}
\newcommand{\upd}{ {\rm d} }

\newcommand{\partialLeft}[1]{\overleftarrow{\partial}_{\!\! #1} }
\newcommand{\partialRight}[1]{\overrightarrow{\partial}_{\!\! #1} }

\newcommand{\xproduct}{\!\times\!}
\newcommand{\sproduct}{\!\cdot\!}

\newcommand{\vect}[1] { {\boldsymbol{#1}} }

\newcommand{\Tr}{\mathrm{Tr}}

\newcommand{\abs}[1]{{\left|#1\right|}}
\newcommand{\avg}[1]{\left\langle#1\right\rangle}
\newcommand{\ket}[1]{| {#1} \rangle}
\newcommand{\bra}[1]{\langle {#1} |}

\newcommand{\sig}{\vect \sigma}

\newcommand{\calA}{\mathcal{A}}
\newcommand{\calB}{\mathcal{B}}

\newcommand{\calD}{\mathcal{D}}
\newcommand{\calE}{\mathcal{E}}

\newcommand{\calG}{\mathcal{G}}
\newcommand{\calH}{\mathcal{H}}

\newcommand{\calM}{\mathcal{M}}
\newcommand{\calN}{\mathcal{N}}
\newcommand{\calO}{\mathcal{O}}
\newcommand{\calP}{\mathcal{P}}

\newcommand{\calR}{\mathcal{R}}
\newcommand{\calS}{\mathcal{S}}
\newcommand{\calT}{\mathcal{T}}
\newcommand{\calU}{\mathcal{U}}

\newcommand{\calX}{\mathcal{X}}
\newcommand{\calY}{\mathcal{Y}}

\newcommand {\Moyal} { \ast }
\newcommand {\MoyalKinetic} { \star }
\def\myTens#1{ {\vbox{\hbox{$\smash {#1}$} \vskip0.3ex \hrule \vskip0.2ex \hrule \vskip-0.68ex }} }

\newcommand{\fD}  { f_{\scriptscriptstyle \rm D} }	

\newcommand{\Ad}[1]  {\calA_{#1}^{\rm (d)}}
\newcommand{\Adp}[1]  {\calA_{#1}^{\rm \prime(d)}}   

\newcommand{\BP} {\calB^{\scriptscriptstyle (\vect p)}}
\newcommand{\BR} {\calB^{\scriptscriptstyle (\vect r)}}
\newcommand{\EP} {\calE^{\scriptscriptstyle (\vect p)}}
\newcommand{\ER} {\calE^{\scriptscriptstyle (\vect r)}}
\newcommand{\ThetaRP} {\myTens \Theta^{\scriptscriptstyle \vect r\! \vect p}}
\newcommand{\BPp} {\calB^{\prime{\scriptscriptstyle (\vect p)}}}
\newcommand{\BRp} {\calB^{\prime{\scriptscriptstyle (\vect r)}}}
\newcommand{\EPp} {\calE^{\prime{\scriptscriptstyle (\vect p)}}}
\newcommand{\ERp} {\calE^{\prime{\scriptscriptstyle (\vect r)}}}
\newcommand{\ThetaRPp} {\myTens \Theta^{\prime{\scriptscriptstyle \vect r\! \vect p}}}

\newcommand{\ORR} {\Omega^{\scriptscriptstyle \vect r \! \vect r}}
\newcommand{\OTR} {\Omega^{\scriptscriptstyle t \vect r}}
\newcommand{\OPP} {\Omega^{\scriptscriptstyle \vect p \! \vect p}}
\newcommand{\OTP} {\Omega^{\scriptscriptstyle t \vect p}}

\newcommand{\OmegaRP} {\myTens \Omega^{\scriptscriptstyle \vect r\! \vect p}}
\newcommand{\OmegaPR} {\myTens \Omega^{\scriptscriptstyle \vect p\! \vect r}}
\newcommand{\OmegaRPp} {\myTens \Omega^{\prime{\scriptscriptstyle \vect r\! \vect p}}}

\begin{document}

\title{Effective Quantum Theories for Transport in Inhomogeneous Systems with Non-trivial Band Structure}
\author{Christian Wickles}
\email[]{christian.wickles@uni-konstanz.de}
\affiliation{Universit\"at Konstanz, Fachbereich Physik, 78457
  Konstanz, Germany}
\author{Wolfgang Belzig}
\email[]{wolfgang.belzig@uni-konstanz.de}
\affiliation{Universit\"at Konstanz, Fachbereich Physik, 78457
  Konstanz, Germany}

\date{\today}

\begin{abstract}
   Starting from a general $N$-band Hamiltonian with weak spatial and temporal variations, we derive a low energy effective theory for transport within one or several overlapping bands. To this end, we use the Wigner representation that allows us to systematically construct the unitary transformation that brings the Hamiltonian into band-diagonal form. We address the issue of gauge invariance and discuss the necessity of using kinetic variables in order to obtain a low energy effective description that is consistent with the original theory. Essentially, our analysis is a semiclassical one and quantum corrections appear as Berry curvatures in addition to quantities that are related to the appearance of persistent currents. We develop a transport framework which is manifestly gauge invariant and it is based on a quantum Boltzman formulation along with suitable definitions of current density operators such that Liouville's theorem is satisfied. Finally, we incorporate the effects of an external electromagnetic field into our theory.
\end{abstract}

\pacs{03.65.Sq,03.65.Vf,72.10.Bg,73.43.-f}


\maketitle

\renewcommand{\sig}{\vect \sigma}
\newcommand{\hdag}{{\scriptstyle \diagup \hspace{-.24cm}} \hbar}

\section{Introduction}
When performing a semiclassical analysis, one naturally encounters Berry phases \cite{Berry1984} and meanwhile, the importance of these so-called geometrical phases in condensed matter physics is beyond question \cite{Mikitik1999,RevModPhysXiao2010,Fuchs2010,Hals2010}. For example, not long ago it was realized that the electric polarizability can be defined in terms of a Berry curvature, for the first time a consistent formulation of this subject \cite{Vanderbilt1993}. Furthermore, research in the field of the anomalous Hall effect (AHE) has shown that the intrinsic contribution is related to a Berry curvature, which is a quantum mechanical property of a perfect crystal \cite{Haldane2004,Sinitsyn2008,NagaosaAHEReview2010}. Also, magnetic monopoles appearing in the definition of a momentum space effective magnetic field give important modifications to universal conductance fluctuations \cite{Hals2010}. Finally, topological interference effects arise from spin Berry phases in single molecular magnets\cite{Leuenberger2001}.

There are different ways to obtain semiclassical transport equations (see Refs. \onlinecite{RevModPhysXiao2010,Fuchs2010} and references therein), like wave-packet analysis \cite{Sundaram1999,Culcer2005}, or the systematic diagonalization method developed by Gosselin and coworkers \cite{Gosselin2007,Gosselin2008,Gosselin2008EPL} which, however, does not include the possibility of time-dependent perturbations.
Furthermore, there are various works treating semiclassical quantum transport equations that incorporate Berry phase effects: the scenario of a general 2-band model is considered by Wong and Tserkovnyak\cite{Wong2011}, and spin-orbit coupled systems \cite{Culcer2006} as well as a non-Abelean gauge-field formulation \cite{Shelankov2010} is investigated. The L\"owdin partitioning, or quasi-degenerate perturbation theory, used in the book of Winkler \cite{WinklerBook2003} to derive effective models for certain bands in spin-orbit coupled semiconductors can also be related to a semiclassical treatment, however, there focus is only put on the Hamiltonian, not on the dynamical variables or other aspects of the system.

In this work, we present a self-contained derivation of the semiclassical dynamics which is based on the Wigner representation \cite{Wigner1932} -- or phase-space representation of quantum mechanics -- which is a natural starting point for a semiclassical analysis. One big advantage is that one can obtain corrections systematically to arbitrary order in $\hbar$. Also, a re-quantization of the effective theory is not necessary, which is a big drawback of the wave-packet analysis which derives a Lagrangian from the equations of motion for the wave-packet center of mass coordinates, and it is not always clear what the canonical conjugate variables are. The relation between canonical and kinetic pairs of conjugate variables, however, emerges naturally in our formalism.
We adopt a 4-component vector notation which allows us to incorporate spatial inhomogeneity as well as temporal variation on an equal basis. In the course of our treatment, we will find how fictitious electric and magnetic fields (real space and its momentum space pendants) appear in effective theories and we complete our work by developing a low energy effective quantum transport theory which is manifestly gauge invariant and consistent with a description in the original frame. Finally, we address the interesting question of how an external electromagnetic field modifies the formalism.

We have several scenarios in mind of applying our formalism to: studying the electron dynamics in the presence of an arbitrary inhomogeneous and time-dependent ferromagnetic exchange field, which exhibits many interesting phenomena \cite{Tatara2004,Tatara2008rev,Wong2009}. One can make various generalizations like adding a spin-orbit coupling term, which would give rise to an even much broader range of new effects such as the anomalous Hall effect, and the latter would be additionally modified by the inhomogeneous magnetization. Studying spin-transport phenomena and/or adding thermal gradients delivers a whole new range of possibilities. \cite{Bauer2012NAT}
Furthermore, there is a recently discovered class of materials called topological insulators, in which the quasi-particle momentum is intimately coupled to the spin-degree of freedom and thus behaves similar to a relativistic Dirac Fermion.\cite{Hsieh2008,QiZhangRMP2011}
Now, adding an additional coupling between spatial and spin degrees-of-freedom leads to a physically rich system already subject to various studies. \cite{Qi2008NAT,Yokoyama2010MagDyn,Nomura2010,Wickles2012STIDW,TserLoss2012}
Considering the effects of mechanical rotation of these systems \cite{MatsuoPRL2011,MatsuoPRB2011} is another interesting application for our formalism.

Approaches that diagonalize band-space in order to obtain an effective low energy description have been performed in the case of quasi-free electrons in ferromagnetic metals with spatiotemporally varying exchange field. There, one can come up with a position-dependent unitary matrix $U(\vect r)$ that diagonalizes spin-space which maps the problem on that of a free particle in an (fictitious) electromagnetic field. \cite{TserkovnyakWong2009}
The converse situation of coupling between spin and momentum as given by the spin-orbit interaction can equally be treated by a momentum-dependent unitary matrix $U(\vect p)$, for example the Foldy-Wouthuysen transformation in the case of the relativistic Dirac equation \cite{FoldyWouthuysen1950}.

It is clear that the full quantum mechanical problem of surface Dirac Fermions coupled to a general inhomogeneous and time-dependent magnetization texture will in general be very complex due to the locking of spin and momentum as well as coupling between spin and spatial degrees of freedom. If one wants to diagonalize spin space, the unitary transformation has to involve the pair of canonical operators $\vect r$ and $\vect p$ which is rather difficult due to non-commutation of $\vect r$ and $\vect p$.
Our approach relies on the fact that, if $\vect r$ and $\vect p$ are classical variables, such a unitary transformation is much more simple to find, and then, we resort to quantum corrections which are of the order $\hbar$. Despite the fact that we are performing formally an expansion in $\hbar$, it does not necessarily need to be restricted to the semiclassical regime. In fact, as pointed out later, our actual expansion parameter might be a different one, depending on the physical system and the regime under investigation. For example, in the case of the Dirac theory, as we will discuss thoroughly in section \ref{sec.DiracEquation}, the actual scale relevant for our expansion is the Compton wavelength $\lambda_c = \frac{\hbar} {m c}$ so the resulting Pauli-Equation still correctly describes the quantum regime for scalar potentials smooth on the scale of $\lambda_c$.

The outline of this work is as follows: In section \ref{sec:diagonalization}, we will introduce the unitary transformation that performs a rotation within band-space such that the Hamiltonian becomes band-diagonal. Since this transformation is not uniquely defined, we will discuss the implications of this additional gauge degree of freedom. This motivates a description in terms of kinetic variables, which leads to the appearance of Berry curvatures, which is discussed in some detail in section \ref{sec.kineticDescription}. We will also investigate how observables change in the course of the diagonalization and we discuss the electronic spectrum as well as energy corrections appearing therein.
Subsequently, in section \ref{sec.EffectiveTheoryEquationsOfMotion} we develop a manifest gauge invariant description of the physical system restricted to a certain band, i.e. we seek a projected theory without the necessity to refer back to the original Hamiltonian. To this end, we find equations of motion for the quasi-probability density, essentially a quantum Boltzmann equation applicable to a non-equilibrium scenario. We also find current densities that obey a conservation law corresponding to Liouville's theorem in classical mechanics. Our quantum mechanical equations of motions are formally similar to the equations of motion for the center-of-mass motion of a wave-packet and, in fact, the latter is just a special case of our formulation.
Finally, in section \ref{sec.HierarchyAndElectromagneticField}, we illustrate how to treat the external electromagnetic field which can be done in the spirit of a hierarchy of effective theories. After concluding the discussion of general systems, in section \ref{sec.DiracEquation} we apply the apparatus developed to the Dirac equation and readily find a relativistic version of the Pauli-Hamiltonian, thereby gaining some interesting insights into the structure of the Dirac equation.
For a concise summary of this work, the reader might want to go to section \ref{sec.DiracEquation} where all central results are being referenced and find immediate application.


\section{The Quest for a Band-diagonal Hamiltonian}
\label{sec:diagonalization}

\begin{figure}
  \centering
  \includegraphics[width=3.5cm]{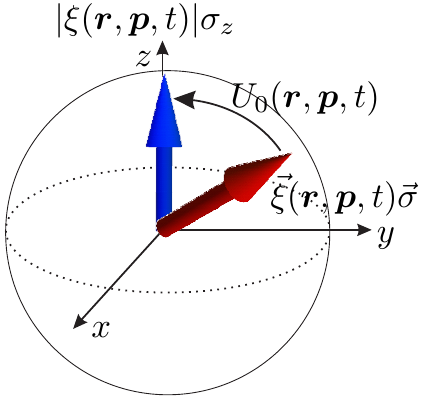}
  \caption{\label{fig:BlochSphere} Illustration of the band-diagonalization scheme: Bloch-sphere representation for a simple 2-band model $\calH = \vec \xi(\vect r, \vect p, t) \cdot \vec \sig = \xi_x \sigma_x + \xi_y \sigma_y + \xi_z \sigma_z$ where the vector is rotated by $U_0(\vect r, \vect p,t)$ such that it aligns along the $z$-axis at each point in phase-space. }
\end{figure}

We now consider a Hamiltonian that consists of $N$ bands and which is almost diagonal in momentum space $\vect p$, but has some additional spatial- and/or temporal variation imprinted on it. In the usual quantum representation, it is expressed in terms of the canonical pair of operators $[\hat r_i, \hat p_j] = i\hbar \delta_{ij}$ and additionally carries the $N \times N$ dimensional matrix structure.
Prominent examples that fall into this category are the Dirac equation to be studied in more detail in section \ref{sec.DiracEquation} as well as the aforementioned system of surface Dirac fermions coupled to a spatially dependent magnetization.

For practical reasons, we resort to a description in terms of single particle Greens functions, or more specifically the inverse thereof,
$\Xi \equiv i\hbar \partial_t \mathbbm 1_N  - \calH(\hat {\vect r}, \hat {\vect p}, t)$.
One reason is that only $\Xi$ represents the complete equation of motion, i.e. all kinetic equations involve this operator, and not $\calH$ alone, for example, we can generally write quantum kinetic equations in the compact form
\begin{align}
\label{eq:GeneralKineticEquation}
\left[ \Xi , \ \calD \right] = 0 \ ,
\end{align}
where $\calD$ represents any dynamical variable like the usual retarded or Keldysh Greens functions, density of states or the density matrix for which (\ref{eq:GeneralKineticEquation}) reduces to the well-known von Neumann equation
\begin{align}
i\hbar \partial_t \rho = \left[ \calH, \rho \right] \ .
\end{align}
The properties of $\calD$ (the type of Greens function, etc.) enter through appropriate boundary conditions in our parameter space. For the example just given, we need to fix our density matrix at some initial time $\rho(t_0) = \rho_0$.

We now transform everything into the Wigner representation \cite{Rammer1986}, so that our Hamiltonian and our observables are a function of the variables $\vect r, \vect p$, $t$ and energy $\epsilon$ so that (see Appendix \ref{ap:WignerRepresentation} for more details)
\begin{align}
\Xi(\vect r, t, \vect p, \epsilon) \equiv \epsilon \mathbbm 1_N  - \calH(\vect r, t, \vect p) \ .
\end{align}
Multiplication of operators is now performed by virtue of the Moyal product \cite{Moyal1949},
\begin{align}
\label{eq:MoyalProduct}
\Moyal &\equiv \exp{\frac{i\hbar}2 \Lambda} \ ,
\end{align}
where the differential operator $\Lambda$ is given by
\begin{align}
\Lambda &= \partialLeft{\vect r} \partialRight{\vect p} - \partialLeft{t} \partialRight{\epsilon} - \partialLeft{\vect p} \partialRight{\vect r} + \partialLeft{\epsilon} \partialRight{t} = -\partialLeft{\vect x} \partialRight{\vect \pi} + \partialLeft{\vect \pi} \partialRight{\vect x}
\end{align}
and we introduced the compact 4-component vector notation $\vect x = (t, \vect r)$ and $\vect \pi = (\epsilon, -\vect p)$. Note that we always use $\vect x$, $\partial_{\vect x} = (\partial_t, \partial_{\vect r})$ and $\calA_{\vect x} = (\calA_t, \calA_{\vect r})$ in contravariant notation whereas we implicitly assume covariant notation for the symbols $\vect \pi$, $\partial_{\vect \pi} = (\partial_\epsilon, -\partial_{\vect p})$ and $\calA_{\vect \pi} = (\calA_\epsilon, -\calA_{\vect p})$. Since contraction will always be between pairs of $\vect x$ and $\vect \pi$, there is no need to indicate covariant vectors.  This provides us a symmetric and compact notation in the following treatment.
The kinetic equation (\ref{eq:GeneralKineticEquation}) is straightforwardly transformed into the Wigner picture:
\begin{align}
\label{eq:GeneralKineticEquationWigner}
\left[ \Xi \stackrel\Moyal, \ \calD \right] = 0 \ .
\end{align}

We are now looking for a unitary matrix $\calU(\vect x, \vect \pi)$ that transforms our initial Hamiltonian $\Xi(\vect x, \vect \pi)$ into a band-diagonal Hamiltonian, denoted by $\bar \Xi(\vect x, \vect \pi)$ in the following, i.e.
\begin{align}
\label{eq:UnitarityCondition}
\calU \Moyal \calU^\dagger = \calU^\dagger \Moyal \calU &= \mathbbm 1 \\
\label{eq:GeneralDiagonalizationTransformation}
\calU \Moyal \Xi \Moyal \calU^\dagger &= \bar\Xi \ .
\end{align}
Let us note that upon this transformation, $\calU \Moyal \epsilon \Moyal \calU^\dagger$ can acquire off-diagonal elements when $\calU$ depends on time, thus requiring the diagonalization of the combination $\Xi = \epsilon \mathbbm{1}_N - \calH$ rather than $\calH$ alone. This is one major difference to previous semiclassical schemes \cite{Gosselin2008}.
For diagonalizations that do not require explicit dependency on the time parameter, treating $\calH$ and $\Xi$ is equivalent, and in the following, we will use the term Hamiltonian equally for both objects.
Note that another advantage of using $\Xi$ instead of $\calH$, which is the form invariance of the kinetic equation (\ref{eq:GeneralKineticEquationWigner}) under unitary transformations.

In the classical limit, the operators $\hat{r}_i$ and $\hat{p}_j$ commute, while in the Wigner representation, the Moyal product (\ref{eq:MoyalProduct}) becomes trivial as $\hbar \rightarrow 0$; after all, in this formulation the Moyal product encodes the non-commutativity of the canonical variables. Then, we essentially have to diagonalize a $N \times N$ matrix $\calH$, whose elements are functions of $\vect r$, $t$ and $\vect p$. We call the unitary matrix associated with this rotation in band-space $U_0(\vect r, \vect p, t)$, so that
\begin{align}
\label{eq:DefinitionU0}
\calU_0 \, \Xi  \, \calU_0^\dagger = \bar\Xi_0 \qquad \Leftrightarrow \qquad \calU_0 \calH \, \calU_0^\dagger = \bar\calH_0 \ ,
\end{align}
and the digonal elements of $\bar\Xi_0$ constitute the classical energies of the $N$ bands described by our initial Hamiltonian.
At any rate, we assume from now on that we know the diagonalization matrix $\calU_0$ analytically. Note that all our matrices parametrically depend on $\vect x$ and $\vect \pi$, so we essentially diagonalize locally at every point in $2\times(3+1)$-dimensional parameter space (see Fig. \ref{fig:BlochSphere} for an illustration), which becomes meaningful in the semiclassical limit, as position $\vect r$, momentum $\vect p$, time $t$ and energy $\epsilon$ are well-defined in this limit.
Also note that $U_0(\vect r, \vect p, t)$ is not uniquely defined which brings up the problem of gauge-invariance as discussed in detail later.

\begin{figure}
  \centering
  \includegraphics[width=4cm]{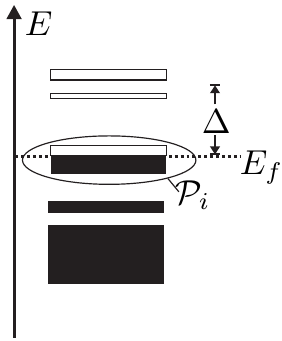}
  \caption{\label{fig:Projection} Illustration of the bands described by the Hamiltonian $\calH$ and $\Delta$ is a typical energy scale for interband distances. The projection operator on the $i$-th band is denoted by $\calP_i$. For low-energy processes around the Fermi energy $E_f$, only one or few overlapping bands are relevant and thus contribute to, e.g transport properties. Corrections due to the  influence of other bands enter as Berry curvatures and are $\propto 1/\Delta$. }
\end{figure}

Furthermore, if the band degrees-of-freedom couple to only either $\vect x$ or $\vect \pi$, then we already found the exact expression $\calU = \calU_0$. For example, treating metallic ferromagnets with inhomogeneous magnetization and neglecting spin-orbit interactions \cite{Wong2009}, it just necessary to only diagonalize $\vect m(\vect r) \sig$, since the energy dispersion is diagonal in spin-space. Then, the effect of $\calU$ is that $\vect p$ acquires an additional vector potential: $\vect p \rightarrow \vect p - i\hbar \calU (\partial_{\vect r} \calU^\dagger)$.

The general situation, where the degrees-of-freedom couple to both spatial and momentum coordinates simultaneously, is much more involved and chances that we can come up with an exact solution for $\calU(\vect x, \vect \pi)$ are slim. Therefore, we adopt a gradient expansion approach where we expand the Moyal Product $\Moyal$ in powers of $\hbar$ and calculate corrections systematically order by order.
The fact that we are performing formally an expansion in $\hbar$ does not necessarily imply that we are restricted to the semiclassical regime. Our actual expansion parameter might be a different one, depending on the physical system and the regime under investigation. Typically, the expansion parameter is the ratio of the energy associated with band-dependent spatiotemporal variations and the interband energy distance $\Delta$, as illustrated in Fig. \ref{fig:Projection} and therefore, the larger the band separations (or weaker coupling), the better the approximation becomes. To emphasize this fact, we will introduce the notation of $\hdag$ in this article to indicate that the expansion is not necessarily a semiclassical one.

We assume that we already solved the zeroth order problem which yields the unitary matrix $\calU_0$, so that now, we can introduce gradient corrections to this matrix by writing
\begin{align}
\calU = \left(\mathbbm{1} + \bar\calU_1 + \bar \calU_2 + \dots \right) \calU_0 \ ,
\end{align}
which describes the matrix that diagonalizes our Hamiltonian $\Xi$ to arbitrary order in $\hdag$. $\bar\calU_1$ is $\propto \hdag$ and will be specified explicitly later.
As a result of the expansion of the Moyal product, we obtain the covariant derivatives
\begin{align}
\label{eq:CovariantDerivative}
\bar\partial_i = \calU_0 \partial_i \calU_0^\dagger = \partial_i - i \calA_i
\end{align}
which acquire the Berry connections
\begin{align}
\label{eq:BerryConnectionMatrices}
\calA_i \equiv i \calU_0 (\partial_i \calU_0^\dagger) \ ,
\end{align}
where $\partial_i$ is a placeholder for any of the possible derivatives $\partial_t, \partial_\epsilon, \partial_{\vect r}$ and $\partial_{\vect p}$.
Berry phases, or often called geometric phases, are omnipresent in modern physics, and describe phases picked up along a trajectory in curved geometries.
Note that $\calA_i$ is a $N \times N$ matrix which is Hermitian
\begin{align}
\calA_i^\dagger = -i (\calU_0 \partial_i \calU_0^\dagger)^\dagger = -i (\partial_i\calU_0) \calU_0^\dagger = i \calU_0 (\partial_i \calU_0^\dagger) = \calA_i \ ,
\end{align}
where the last step is due to $ 0 = \partial_i (\calU_0 \calU_0^\dagger) = (\partial_i \calU_0) \calU_0^\dagger + \calU_0 (\partial_i \calU_0^\dagger)$
and the diagonal elements of $\calA_i$ describe the usual Berry connections arising from transport within a certain band. The off-diagonal elements mix contributions from two bands and thus describe effects due to inter-band coupling during transport in a certain band. As we will see later, these inter-band transitions will give rise to important corrections  for example the correction to the energy that appears to leading order in $\hdag$.
Transforming higher order derivatives is straightforward, except for the additional freedom of exchanging partial derivatives, which leads to a general relation between Berry connection matrices $\calA_i$. Starting from
\begin{align}
\calU_0 \partial_j \partial_i \calU_0^\dagger = \bar\partial_j \bar\partial_i = (\partial_j - i \calA_j) (\partial_i - i \calA_i) \ ,
\end{align}
and likewise,
\begin{align}
\calU_0 \partial_i \partial_j \calU_0^\dagger = \bar\partial_i \bar\partial_j = (\partial_i - i \calA_i) (\partial_j - i \calA_j)  \ ,
\end{align}
and, if we assume symmetry of second derivatives $\partial_j \partial_i = \partial_i \partial_j$, the following relation should hold
\begin{align}
\label{eq:BerryConnectionRelation}
\partial_i \calA_j - \partial_j \calA_i = i\left[ \calA_i, \calA_j \right ] \ .
\end{align}
In fact, this identity can be directly shown by using the definition of $\calA$ (\ref{eq:BerryConnectionMatrices}) and the exchange of partial derivatives.

Now we have everything at hand to systematically calculate $\bar\Xi = \bar\Xi_0 + \bar\Xi_1 + \bar\Xi_2 \dots$ to arbitrary order in $\hdag$
and to first order, we explicitly obtain the following result for the transformed expression,
\begin{widetext}
\begin{align}
\label{eq:TransformationRelationHU}
\bar \Xi_1 = \bar \calU_1 \bar\Xi_0 + \bar\Xi_0 \bar\calU_1^\dagger - \frac{\hbar}{2} \left\{ \calA_{\vect \pi}, \partial_{\vect x} \bar\Xi_0 \right\} + \frac{\hbar}{2} \left\{ \calA_{\vect x} , \partial_{\vect \pi} \bar\Xi_0\right\} + \frac{i\hbar}2 \left( \calA_{\vect \pi} \bar\Xi_0 \calA_{\vect x} - \calA_{\vect x} \bar\Xi_0 \calA_{\vect \pi} \right) \ .
\end{align}
\end{widetext}
We obtain an additional constraint for $\bar\calU_n$ from the condition of unitarity (\ref{eq:UnitarityCondition}), which to first order in $\hdag$ reads
\begin{align}
\label{eq:CalU1Condition}
\bar \calU_1 + \bar\calU_1^\dagger + \frac{i\hbar}2 \left[ \calA_{\vect \pi}, \calA_{\vect x} \right] \stackrel{!}{=} 0 \ ,
\end{align}
and is obtained by substituting $\mathbbm 1$ for $\bar\Xi_0$ into the transformation relation Eq. (\ref{eq:TransformationRelationHU}). Since this relation fixes the Hermitian part of $\bar\calU_1$, we can make the ansatz
\begin{align}
\bar\calU_1 &= -\frac{i\hbar}4 \left[ \calA_{\vect \pi}, \calA_{\vect x} \right] + \calY_1 \nonumber \ , 
\end{align}
with $\calY_1= -\calY_1^\dagger$ assumed to be antihermitian, such that condition (\ref{eq:CalU1Condition}) is satisfied.
Plugging $\bar \calU_1$ back into Eq. (\ref{eq:TransformationRelationHU}) yields
\begin{widetext}
\begin{align}
\label{eq:TransformationRelationXi}
\bar\Xi_1 = \left[\calY_1, \bar\Xi_0 \right] - \frac{\hbar}{2} \left\{ \calA_{\vect \pi} , \left[\partial_{\vect x} - \tfrac i2 \calA_{\vect x}, \bar\Xi_0\right] \right\} + \frac{\hbar}{2} \left\{ \calA_{\vect x} , \left[\partial_{\vect \pi} - \tfrac i2 \calA_{\vect \pi}, \bar\Xi_0\right]  \right\} + O(\hdag^2) \ .
\end{align}
\end{widetext}
We observe that $\left[\calY_1, \bar\Xi_0 \right]$ is completely off-diagonal and $\calY_1$ is well suited to absorb the off-diagonal part of the last two terms on the right-hand-side of Eq. (\ref{eq:TransformationRelationXi}), in the following denoted as $\calR_{\rm o}$. After all, the objective is to diagonalize $\bar\Xi$, so we want the off-diagonal part of $\bar\Xi_1$ to vanish, which is achieved by the condition
\begin{align}
\label{eq:DefinitionY1}
\left[\calY_1, \bar\Xi_0 \right] &= \calR_{\rm o} \ .
\end{align}
The solution can be readily given as
\begin{align}
\label{eq:Y1Explicit}
\Rightarrow (\calY_1)_{ij} &=\frac{\left( \calR_{\rm o} \right)_{ij}}{\epsilon_i - \epsilon_j}
\end{align}
in the case of a completely diagonalized $\left(\bar\calH_0\right)_{ij} = \epsilon_i \delta_{ij}$ and which is antihermitian as desired. Here we explicitly see that corrections due to $\calY$ are inversely proportional to the separation between bands, and thus are of the same order as corrections due to Berry phases. In fact, the commutator on the left side with the Hamiltonian $\bar\calH_0$ is exactly what also appears in the von Neumann equation suggesting further that this term describes corrections due to inter-band dynamics.

We remark that the difference between the diagonalization of $\Xi$ and $\calH$ is merely the modification of $\calY_1$ by that it acquires an additional contribution from the off-diagonal elements of the Berry connection $\calA_t$. However, this term is crucial when our transformation $\calU$ is time-dependent and only then will the formalism yield consistent results.
Note that the definition (\ref{eq:DefinitionY1}) does not fix the imaginary diagonal part of $\calY_1$, but without loss of generality, we can set this part to zero. A non-zero imaginary diagonal part corresponds to the linear order expansion of the gauge phase factors $e^{i\chi(\vect x, \vect \pi)}$ (see (\ref{eq:GaugeFactor})). Essentially, the freedom of choice here can be reduced to the problem of gauge invariance, to be discussed in the next section.

Yet, we also want to capture the situation of degenerate or overlapping bands, so the desired form of $\bar\Xi$ is in general block-diagonal. To formally express this matter, we introduce the projectors $\calP_i$ that define the bands (see Fig. \ref{fig:Projection}) and that, in the end, we want use for our effective theory.
Of course, the set of $\calP_i$ has to be specified together with $\calU_0$, since the unitary transformation has a freedom of how we distribute the bands amongst the entries of our matrix. For example, a natural choice would be to sort the bands with respect to their energies.
Then, introducing the projected Berry connection $\Ad{\vect \pi} \equiv \calP_{\rm d} \calA_{\vect \pi} \calP_{\rm d} \equiv \sum_i \calP_i \calA_{\vect \pi} \calP_i$, we can write for the Hamiltonian that is diagonalized up to first order in $\hdag$,
\begin{widetext}
\begin{align}
\label{eq:effectiveHamiltonian}
\bar\Xi = \bar \Xi_0 - \hbar \Ad{\vect \pi} \partial_{\vect x} \bar\Xi_0 + \hbar \Ad{\vect x} \partial_{\vect \pi} \bar\Xi_0 + \frac{i\hbar}4 \calP_{\rm d} \left(\left\{\calA_{\vect \pi}, \left[\calA_{\vect x}, \bar\Xi_0\right]\right\}  - \left\{\calA_{\vect x}, \left[\calA_{\vect \pi}, \bar\Xi_0\right] \right\}  \right) \calP_{\rm d} \ . 
\end{align}
\end{widetext}
As compared to expression (\ref{eq:TransformationRelationXi}), we can drop two anti-commutators, since after truncation, $\Ad{}$ and $\bar\Xi_0$ commute. As discussed more thoroughly below, this is even the case when either $\bar\Xi_0$ is block-diagonal and/or the projected Berry connections are non-Abelian.

Apart from the obvious term $\bar\Xi_0 = \epsilon - \bar\calH_0$ which can be understood as the classical energy, the last term of (\ref{eq:effectiveHamiltonian}) can be thought of as correction to the energy due to inter-band transitions and corresponds to the energy associated with persistent circulating currents, as for example the magnetic Zeeman energy in the case of the Dirac equation (see discussion in section \ref{sec.DiracEquation}). The two middle terms in (\ref{eq:effectiveHamiltonian}) appear to be Berry phase corrections to this energy, however, they are not unique in the sense that they depend on the specific form of $U_0$. The meaning will become more apparent later, but before, we address the question of gauge invariance of the effective Hamiltonian (\ref{eq:effectiveHamiltonian}).

\subsection{Gauge Invariance}
\label{sec.GaugeInvariance}
We mentioned previously that there is an additional degree of freedom in the choice of unitary transformations $\calU_0$ and $\calU'_0$ which all yield the same diagonal Hamiltonian, $\bar\Xi = \calU_0 \Xi \calU_0^\dagger = \calU'_0 \Xi \calU_0'^\dagger$. These different unitary transformations are related by local phase factors in $2\times(3+1)$- dimensional parameter space, so that we can formally connect two unitary transformations $\calU$ and $\calU'$ by the (block-)diagonal $N \times N$ phase matrix
\begin{align}
\label{eq:GaugeFactor}
\Phi(\vect x, \vect \pi) = \begin{pmatrix} e^{i\chi_1({\vect x, \vect \pi})} & 0 & 0 &  \\ 0  & e^{i\chi_2({\vect x, \vect \pi})} & 0 & \dots &  \\ 0 & 0 & e^{i\chi_3({\vect x, \vect \pi})} & \\  & \vdots &  & \ddots \end{pmatrix} \ ,
\end{align}
where $\chi_n({\vect x, \vect \pi})$ are arbitrary functions in our parameter space, which are appropriately termed gauge fields. In fact, $\calU_0' = \Phi \calU_0$ describes a gauge transformation and can be thought of as a local phase transformation in an extended phase space which includes time and energy. This gauge transformation changes the Berry connection matrices according to
\begin{align}
\calA'_k = i \calU_0' \partial_k \calU_0'^\dagger = \Phi \calA_k \Phi^\dagger + i \Phi \partial_k \Phi^\dagger = \Phi \calA_k \Phi^\dagger + \calX_k \ ,
\end{align}
where the field $\calX_k = i \Phi \partial_k \Phi^\dagger = \partial_k {\rm diag}(\chi_1, \chi_2, \dots, \chi_N) $ is a (block-)diagonal matrix containing the partial derivatives of the phases. $\Phi \calA_k \Phi^\dagger$ modifies the off-diagonal elements by giving them additional phase-factors, while the change in diagonal elements is due to $\calX_k$. In principle, we have to distinguish three different cases here, the first one being that $\bar\Xi_0$ is completely diagonalized with well-separated bands which corresponds to the situation just described.
However, when we keep part of $\bar\Xi_0$ block-diagonal because bands are overlapping or degenerate and we can distinguish the bands, i.e. they have a physical meaning that we want to retain (for example we have spin-degenerate bands but want to describe spin-dependent physics) then we have only a ${\rm U}(1)$ gauge freedom within this sub-block. Third and last, if we have $M$ degenerate bands, i.e. there is a $M$-dimensional sub block in $\bar\Xi_0$ that is proportional to the unit matrix, and furthermore, we cannot or do not want to distinguish between the degenerate bands, we have the additional degree of freedom to rotate within this degenerate space giving us an additional ${\rm SU}(M)$ gauge invariance. Contrary to the first two cases, this last one describes a situation with the effective description of this $M$-dimensional subspace being a non-Abelian gauge theory with the symmetry group ${\rm U}(1) \times {\rm SU}(M)$ and consequently, $\calX_k$ constitutes a non-Abelian field. We do not differentiate between these cases explicitly in the following because they are straightforwardly treated in our formulas, thus requiring us only to comment in situations where special care is required.

According to the preceding discussion, the projected Berry connection matrix is only modified due to $\calX$,
\begin{align}
\Adp{\vect x} = \calP_{\rm d} \calA'_{\vect x} \calP_{\rm d} = \Ad{\vect x} + \calX_{\vect x} \ ,
\end{align}
and the alternative transformation due to $\calU_0'$ leads to the Hamiltonian
\begin{widetext}
\begin{align}
\bar\Xi' &= \calU' \Moyal \Xi \Moyal \calU'^\dagger \nonumber\\
&= \bar \Xi_0 - \frac{\hbar}{2} \left\{ \Adp{\vect \pi}, \partial_{\vect x} \bar\Xi_0 \right\} + \frac{\hbar}{2} \left\{ \Adp{\vect x}, \partial_{\vect \pi} \bar\Xi_0 \right\}
  + \frac{i\hbar}4 \calP_{\rm d} \left(\left\{\calA_{\vect \pi}, \left[\calA_{\vect x}, \bar\Xi_0\right]\right\}  - \left\{\calA_{\vect x}, \left[\calA_{\vect \pi}, \bar\Xi_0\right] \right\}  \right) \calP_{\rm d} +  O(\hdag^2) \ .
\end{align}
\end{widetext}
The last term, representing the inter-band transition corrections to the energy, does not change, since the additional terms due to $\calX_{\vect x}$, $\calX_{\vect \pi}$ are projected out by $\calP_{\rm d}$ and thus, as a consequence, are absorbed by $\calY_1'$ {\it viz.} $\bar\calU_1'$ which of course does not need to coincide with $\bar\calU_1$.

However, the other two terms linear in $\hdag$ do explicitly depend on the gauge, and therefore change the effective Hamiltonian. Clearly, this shows that the effective Hamiltonian alone is an incomplete description as it directly depends on this additional degree of freedom. Therefore, in order to make any sense out of this, we have to identify our physical observables, because in the end, the physical results derived from our effective theory should not depend on a specific gauge.

\section{Canonical versus Kinetic Variables and Gauge Invariant Description}
\label{sec.kineticDescription}

The previous section showed us that there is still an ingredient missing in our effective theories. In order to investigate this matter, let us study the dynamics of our system, and construct the effective theory such that the results obtained within this description are consistent with what one would obtain in the original frame.
The question is now, whether one can find a manifest gauge invariant formulation and, as we will explain in the following, it is indeed possible.

\subsection{Parameter transformation to kinetic variables}
Let us consider the operator $\calS$ that describes some physical observable of our system and, in performing the rotation that brings our Hamiltonian $\calH$ into diagonal form, it transforms our observable $\calS$ along with it. The observable in the rotated frame $\bar\calS$ is then related to the original operator $\calS$ by virtue of relation (\ref{eq:TransformationRelation}).
We now consider the projected system, i.e. we have in mind to develop an effective, yet exhaustive description of the physics taking place within a certain band that is sufficiently well separated from all other bands in order to treat this band independently to a good approximation.
For the moment, let us assume that our observable $\calS$ is band-diagonal, i.e. it is a scalar function of $\vect x$ and $\vect \pi$, so that the last term in relation (\ref{eq:TransformationRelation}) vanishes and we have
\begin{align}
\label{eq:TransformationRelationKinetic}
\bar\calS = \bar \calS_0 - \frac{\hbar}{2} \left\{ \Ad{\vect \pi} , \partial_{\vect x} \bar\calS_0 \right\} + \frac{\hbar}{2} \left\{ \Ad{\vect x}, \partial_{\vect \pi} \bar\calS_0 \right\} + O(\hdag^2) \ ,
\end{align}
where $\bar \calS_0 = \calU_0 \calS \calU_0^\dagger$.
Later, we will lift this restriction and consider a matrix $\calS$ with general structure in band space, so that we will also get this additional term, giving rise to important contributions. However for the moment, (\ref{eq:TransformationRelationKinetic}) is nothing but a Taylor expansion of $\bar\calS_0$ to first order in $\hbar$,
\begin{align}
\bar \calS = \bar \calS_0(\vect x - \hbar \Ad{\vect \pi}, \vect \pi + \hbar \Ad{\vect x}) + O(\hdag^2) \ ,
\end{align}
suggesting the parameter transformation to band projected kinetic variables $\vect X$ and $\vect \Pi$
\begin{align}
\label{eq:CanonicToKineticPair}
\vect X^{\rm (d)} &= \vect x - \hbar \Ad{\vect \pi} \\
\vect \Pi^{\rm (d)} &= \vect \pi + \hbar \Ad{\vect x} \ ,
\end{align}
so that we can write
\begin{align}
\bar \calS = \bar \calS_0(\vect X, \vect \Pi) + O(\hdag^2) \ .
\end{align}
We note that to leading order in $\hdag$, we can equally write $\calA(\vect x, \vect \pi) = \calA(\vect X, \vect \Pi) + O(\hdag)$, since Berry connection terms are already linear in $\hdag$.

$\vect P = \vect p - \hbar \calA_{\vect r}$ is of course well known in the Hamilton formulation of particles in an electromagnetic field. In an analogous way, the position $\vect R = \vect r + \hbar \calA_{\vect p}$ acquires an additional Berry connection with its corresponding Berry curvature, or momentum space magnetic field that gives rise to the so called anomalous velocity term. Furthermore, $E = \epsilon + \hbar \calA_t$ attains a contribution which gives rise to an electromotive force and appears in the form of an effective electric field (see for example Ref. \onlinecite{Barnes2007}). The same is true for the electromagnetic field, where the electric field can be also rewritten in terms of a time-dependent phase, effectively changing the gauge. Finally, for reasons of symmetry, one would also have $T = t - \hbar \calA_\epsilon$ but, at least in non-interacting Hamiltonian systems, $\calA_\epsilon$ is zero, since the energy-dependency in $\Xi$ is trivial. However, the situation is different if one considers an interacting system and uses an effective non-interacting quasi-particle description, because the self-energy that includes these interaction in general carries a non-trivial energy-dependence inherited by the single-particle greens function and thus by $\Xi$. \cite{Balents2008}

Let us have a look at the commutator relations between the kinetic variables, \cite{Gosselin2007,Gosselin2008EPL}
\begin{align}
\left[ \vect R_i^{\rm (d)} \stackrel{\Moyal}{,} \vect P_j^{\rm (d)} \right] &= i\hbar \left(\delta_{ij} + \ThetaRP_{ji} \right) \nonumber\\
\label{eq:KineticCommutatorRelations}
\left[ \vect R_i^{\rm (d)} \stackrel{\Moyal}{,} \vect R_j^{\rm (d)} \right] &= i\hbar \epsilon_{ijk} \BP_k \\
\left[ \vect P_i^{\rm (d)} \stackrel{\Moyal}{,} \vect P_j^{\rm (d)} \right] &= i\hbar \epsilon_{ijk} \BR_k \ , \nonumber
\end{align}
where we introduced
\begin{align}
\label{eq:EffectiveFieldBR}
\vect \BR &= \hbar \left( \partial_{\vect r} \times \Ad{\vect r} \right) - i\hbar \left(\Ad{\vect r} \times \Ad{\vect r} \right) \\
\label{eq:EffectiveFieldBP}
\vect \BP &= \hbar \left( \partial_{\vect p} \times \Ad{\vect p} \right) - i\hbar \left(\Ad{\vect p} \times \Ad{\vect p} \right) \ ,
\end{align}
which can be considered as a generalized magnetic field in real space and reciprocal (or momentum) space. Such non-Abelian Berry curvatures have been treated in Ref. \onlinecite{Shelankov2010}.
Furthermore,
\begin{align}
\label{eq:DefinitionThetaRP}
\ThetaRP_{ij} = \hbar \left( \frac{\partial \Ad{\vect p_j}}{\partial \vect r_i} - \frac{\partial \Ad{\vect r_i}}{\partial \vect p_j} - i \left[ \Ad{\vect r_i} , \Ad{\vect p_j} \right]\right)
\end{align}
As we will also encounter later, the dimensionless tensor $\ThetaRP$ describes the change in the metric of the phase-space due to the parameter transformation from canonical to kinetic variables (\ref{eq:CanonicToKineticPair}). We can make this more apparent by relating it to the change in differentials
\begin{align}
\upd R_i &= \left(\delta_{ij} + \hbar \partial_{r_j} \calA_{p_i} \right) \upd r_j \\
\upd P_i &= \left(\delta_{ij} - \hbar \partial_{p_j} \calA_{r_i} \right) \upd p_j \ ,
\end{align}
so that
\begin{align}
\upd \vect R \sproduct \upd \vect P = \upd \vect r \left(\myTens 1 + \ThetaRP \right) \upd \vect p + O(\hdag^2) \ .
\end{align}

A more compact way to write these Berry curvatures is to use the covariant derivative (\ref{eq:CovariantDerivative}), projected onto our band-diagonal space $\bar\partial_{\vect x}^{\rm (d)} = \calP_{\rm d} \bar\partial_{\vect x} \calP_{\rm d}$, for example
\begin{align}
\ThetaRP_{ij} = \hbar \left( \bar \partial_{\vect t_i}^{\rm (d)} \Ad{\vect p_j} - \bar \partial_{\vect p_j}^{\rm (d)} \Ad{\vect r_i} \right)  \ .
\end{align}
It is well known that the Berry curvatures are invariant with respect to gauge transformations, and the commutator is essential as it provides the full ${\rm SU}(M)$ gauge invariance in the non-Abelian case.

In accordance with the effective magnetic fields (\ref{eq:EffectiveFieldBR}) and (\ref{eq:EffectiveFieldBP}), we introduced the effective electric fields
\begin{align}
\label{eq:EffectiveFieldER}
\vect \ER &= \hbar \left( \partial_{\vect r}\Ad{t} - \partial_t \Ad{\vect r} - i\left[ \Ad{\vect r}, \Ad{t} \right] \right) \\
\label{eq:EffectiveFieldEP}
\vect \EP &= \hbar \left( \partial_{\vect p}\Ad{t} - \partial_t \Ad{\vect p} - i\left[ \Ad{\vect p}, \Ad{t} \right] \right) \ ,
\end{align}
which shows us indeed that $\calA_t$ appears in the role of a generalized electric potential, however, it can also depend on momentum $\vect P$.
In the Abelian case (for non-Abelian fields, it works if we take $\Tr_{M} \vect \calE$ and $\Tr_{M} \vect \calB$ or when we take the covariant derivatives (\ref{eq:CovariantDerivative}) along with the full matrix structure of the Berry connections), these fictitious fields obey homogeneous Maxwell equations
\begin{align}
\label{eq:EffectiveMaxwellEquations}
\vect \partial_{\vect r/\vect p} \sproduct \vect \calB^{\scriptscriptstyle(\vect r/\vect p)}  &= 0 \nonumber \\
\vect \partial_{\vect r/\vect p} \xproduct \vect \calE^{\scriptscriptstyle(\vect r/\vect p)} + \partial_t \vect \calB^{\scriptscriptstyle(\vect r/\vect p)} &= 0 \ ,
\end{align}
however, in order to determine these fields independently as in classical electrodynamics, one would need two additional inhomogeneous equations containing (effective) source terms as inhomogeneities.
Note that in general, $\vect \calB^{\scriptscriptstyle(\vect r/\vect p)}$ and $\vect \calE^{\scriptscriptstyle(\vect r/\vect p)}$ depend on $\vect r$ and $\vect p$ simultaneously. As in (\ref{eq:KineticCommutatorRelations}), the effective magnetic fields can be also defined in terms of commutator relations
\begin{align}
\vect \ER &= \frac1{i\hbar} \left[ E^{\rm (d)} \stackrel{\Moyal}{,} \vect P^{\rm (d)} \right] \nonumber \\
\label{eq:KineticCommutatorRelationsEField}
\vect \EP &= -\frac1{i\hbar} \left[ E^{\rm (d)} \stackrel{\Moyal}{,} \vect R^{\rm (d)} \right] \ .
\end{align}

There exists a sum rule for the fictitious fields
\begin{align}
\label{eq:SumRule}
\Tr_{N} \vect \BP &= 0 \ , \qquad \Tr_{N} \vect \BR = 0 \ , \qquad \Tr_{N} \ThetaRP = 0 \ ,\nonumber \\
\Tr_{N} \vect \EP &= 0 \ , \qquad \Tr_{N} \vect \ER = 0 \ ,
\end{align}
which can be found by taking the trace over all bands (we denote this sum over bands as $\Tr_{N}$ here and throughout this work) and making use of the identity (\ref{eq:BerryConnectionRelation}). This means that all the effective forces for each band balance each other in total, or in other words, if all bands are completely filled (and thus, the density operator $\rho$ is proportional to the unit matrix $\mathbbm 1_N$), the system does not experience any net force.

The basis of all calculations within the Wigner representation of quantum theory is the Moyal bracket $\left[ \bar \calS \stackrel{\Moyal}{,} \bar \calT \right]$ between two operators $\bar\calS$ and $\bar\calT$. Now, we want to rewrite this in terms of kinetic variables only, which is achieved by transforming the derivatives in the Moyal product to act on kinetic variables.
For concise notation, we introduce the tensor of Berry curvatures as ($\alpha$, $\beta$ denote indices with respect to $(t, \vect r, \vect p)$)
\begin{align}
\label{eq:ThetaTensorDefinition}
\myTens \Theta_{\alpha,\beta} = \hbar \left( \partial_\alpha \Ad\beta - \partial_\beta \Ad\alpha - i \left[ \Ad\alpha , \Ad\beta \right] \right) \ ,
\end{align}
or explicitly in terms of the fictitious fields used previously
\begin{align}
\label{eq:ThetaTensorDefinitionMatrixForm}
\myTens \Theta = \begin{pmatrix} 0 & - \vect \ER &  -\vect \EP \\ \vect \ER & \epsilon_{ijk} \BR_k & \ThetaRP \\ \vect \EP &  -(\ThetaRP)^T &  \epsilon_{ijk} \BP_k \end{pmatrix} \ ,
\end{align}
where we recognize the top-left part of $\myTens \Theta$ as being essentially the electromagnetic field tensor.

Neglecting terms of order $O(\hdag^2)$, we find the explicit form of the Moyal product after the transformation to kinetic variables
\begin{widetext}
\begin{align}
\label{eq:MoyalTransform}
\Moyal \rightarrow \exp\left\{ \frac{i\hbar}2 \left(-\partialLeft{\vect X} \partialRight{\vect \Pi} + \partialLeft{\vect \Pi} \partialRight{\vect X}\right) + \frac{i\hbar}2 \begin{pmatrix} \partialLeft{\vect \Pi} & \partialLeft{\vect R} \end{pmatrix} \myTens\Theta \begin{pmatrix} \partialRight{\vect \Pi} \\ \partialRight{\vect R} \end{pmatrix} \right\} \ .
\end{align}
\end{widetext}
This is a central result of this section, since it shows that expressing all quantities in terms of kinetic variables allows us to deal solely with manifest gauge invariant expressions. Essentially, the bottom line of this parameter transformation is that it changes the metric of the Moyal product by the appearance of the Berry curvatures $\myTens\Theta$. Immediate consequences of result (\ref{eq:MoyalTransform}) are the equations of motions to be discussed in detail in section \ref{sec.EffectiveTheoryEquationsOfMotion}.


\subsection{Transformation of general operators}
Let us briefly comment on observables with non-trivial band matrix structure $\calO(\vect x, \vect \pi)$, and their transformation into the rotated frame, which is performed analogously to the transformation (\ref{eq:TransformationRelationXi}),
\begin{widetext}
\begin{align}
\label{eq:TransformationRelation}
\bar\calO = \calU \Moyal \calO \Moyal \calU^\dagger = \bar \calO_0 + \left[\calY_1, \bar\calO_0 \right] - \frac{\hbar}{2} \left\{ \calA_{\vect \pi} , \left[\partial_{\vect x} - \tfrac i2 \calA_{\vect x}, \bar\calO_0\right] \right\} + \frac{\hbar}{2} \left\{ \calA_{\vect x} , \left[\partial_{\vect \pi} - \tfrac i2 \calA_{\vect \pi}, \bar\calO_0\right]  \right\} + O(\hdag^2) \ ,
\end{align}
and the back transformation is given by
\begin{align}
\label{eq:BackTransformationRelation}
\calU_0 \calO \calU_0^\dagger = \calU_0 (\calU^\dagger\Moyal\bar\calO\Moyal \calU) \calU_0^\dagger=  \bar \calO - \hbar \left[\calY_1, \bar\calO  \right] + \frac{\hbar}{2} \left\{ \calA_{\vect \pi} , \left[\partial_{\vect x} - \tfrac i2 \calA_{\vect x}, \bar\calO \right] \right\}
  - \frac{\hbar}{2} \left\{ \calA_{\vect x} , \left[\partial_{\vect \pi} - \tfrac i2 \calA_{\vect \pi}, \bar\calO \right]  \right\} + O(\hdag^2)  \ ,
\end{align}
\end{widetext}
which can be readily checked by plugging Eqn (\ref{eq:TransformationRelation}) into the back transformation (\ref{eq:BackTransformationRelation}) and dropping terms of order $\hdag^2$.
Now, treating the diagonal and off-diagonal part of $\bar\calO_0 = \calU_0 \calO \calU_0^\dagger = \bar\calO^{\rm (d)}_0 + \bar\calO^{\rm (o)}_0$ separately, we find
\begin{widetext}
\begin{align}
\label{eq:TransformObservableDiagPart}
\bar \calO^{\rm (d)}_0(\vect X, \vect \Pi) + \frac{i\hbar}{4} \calP_{\rm d} \left( \left\{ \calA_{\vect \pi} , \left[ \calA_{\vect x}, \bar\calO^{\rm (d)}_0\right] \right\} - \frac{i\hbar}{4} \left\{ \calA_{\vect x} , \left[ \calA_{\vect \pi}, \bar\calO^{\rm (d)}_0\right]  \right\} \right) \calP_{\rm d} + O(\hdag^2) \ ,
\end{align}
and the contribution arising from the off-diagonal part of $\bar\calO^{\rm (o)}_0$,
\begin{align}
\label{eq:TransformObservableOffDiagPart}
\calP_{\rm d} \left( \left[\calY_1, \bar\calO^{\rm (o)}_0 \right] - \frac{\hbar}{2} \left\{ \calA_{\vect \pi} , \left[\partial_{\vect x} - \tfrac i2 \calA_{\vect x}, \bar\calO^{\rm (o)}_0\right] \right\} + \frac{\hbar}{2} \left\{ \calA_{\vect x} , \left[\partial_{\vect \pi} - \tfrac i2 \calA_{\vect \pi}, \bar\calO^{\rm (o)}_0\right]  \right\} \right) \calP_{\rm d} + O(\hdag^2) \ ,
\end{align}
\end{widetext}
which are both independently gauge invariant. While the gauge invariance of the former is straightforward to show, the later requires more work and we have to take into account that $\calY_1$ is modified under a gauge transformation as
\begin{align}
\calY_1 \rightarrow \calY_1 - \frac{i\hbar}4 \left( \left\{ \chi_{\vect x} , \calA_{\vect \pi} \right\} - \left\{ \chi_{\vect \pi} , \calA_{\vect x} \right\} \right) \ ,
\end{align}
along with $\partial_{\alpha} \bar\calO \rightarrow \partial_\alpha \bar\calO + i \left[ \chi_\alpha , \bar\calO \right]$ and $\calA_\alpha \rightarrow \calA_\alpha + \chi_\alpha$.

\subsection{Expectation values in the rotated frame}
Let us now go back to the initial question of the dynamics of our system within the effective theory by studying the expectation values of physical observables, which can be obtained in the Wigner representation by the integration over the complete phase space,
\begin{align}
\avg{\calS} = \int \upd^\dm r \int \frac{\upd^\dm p}{(2\pi\hbar)^\dm} \ \Tr_{N}\left\{ \rho(\vect x, \vect \pi) \Moyal \calS(\vect x, \vect \pi) \right\} \ ,
\end{align}
and the trace is with respect to the matrix structure. Note that the factor $1/(2\pi\hbar)^\dm = 1/h^\dm$ describes the proper quantization of the phase space volume and thus is directly obtained by transforming quantum averages into the Wigner representation.

If we assume the integration over the whole phase space to be unbounded and any surface contribution from the integrand at infinity to vanish, we can perform partial integrations to show that all the partial derivatives in the Moyal product $\Moyal$ cancel each other, so that we can equally write
\begin{align}
\avg{\calS} = \int \upd^\dm r \int \frac{\upd^\dm p}{(2\pi\hbar)^\dm} \ \Tr_{N}\left\{ \rho(\vect x, \vect \pi) \calS(\vect x, \vect \pi) \right\} \ .
\end{align}

By using the cyclic property of the trace and by partial integration we can easily show that
\begin{align}
\label{eq:ExpectationValueInRotatedFrame}
\avg{\calO} &= \int  \frac{\upd^\dm r \ \upd^\dm p}{(2\pi\hbar)^\dm} \ \Tr_{N}\left\{ \calU \Moyal \calO \Moyal \calU^\dagger \right\} = \avg{\bar\calO} \ ,
\end{align}
and we can write
\begin{align}
\avg{\calS} = \avg{\bar\calS} = \int \upd^\dm r \int \frac{\upd^\dm p}{(2\pi\hbar)^\dm} \ \Tr_{N}\left\{ \bar\rho(\vect x, \vect \pi) \bar\calS(\vect x, \vect \pi) \right\} \ .
\end{align}

In equilibrium, the density operator is given by
\begin{align}
\rho(\vect x, \vect \pi) =  \fD(\epsilon) \, \delta(\bar \Xi) =  \fD(\epsilon) \  \delta(\epsilon - \bar\calH(\vect x, \vect \pi)) \ ,
\end{align}
or, if we do not want our results to be energy resolved, we use directly
\begin{align}
\rho(\vect x, \vect p) = \int \upd\epsilon \ \fD(\epsilon) \delta(\epsilon - \bar\calH(\vect x, \vect \pi)) \ ,
\end{align}
which gives us the density in phase space and, as we will discuss later however, it is to be interpreted as a quasi-probability. An explicitly time-dependent Hamiltonian $\bar\calH(\vect x, \vect \pi)$ has to be treated using (\ref{eq:BoltzmannEquation}) instead.
If a band is completely filled, it becomes $1$ at the diagonal element corresponding to that band. If energies are well separated, and we assume excitations localized in energy space, we can assume any off-diagonal elements in the density operator to vanish. In fact, those off-diagonal entries correspond to coherent excitations that are split amongst several bands. If the band splitting is sufficiently large, this leads to rapid oscillations subject to decoherence.
In the end, the projection operation defined by $\calP_{\rm d}$ is essentially enforced by the diagonal representation of the Hamiltonian $\bar \Xi$, and by the density matrix $\rho$ which gives us only those states that have coherences within bands (or degenerate/overlapping bands so that, again, the energy argument applies). In particular, this is certainly true for low energy transport, where physics takes place only in the vicinity of the Fermi level.

In order to proceed, let us transform the integration variables to kinetic ones, and in doing so, we also have to take into account how the volume element in phase space changes, which is given by the determinant of the Jacobian
\begin{align}
D^{-1} &\equiv \det \frac{\partial(\vect R, \vect P)}{\partial(\vect r, \vect p)}
= \det \left( \myTens1 + \ThetaRP \right) \nonumber \\
&= 1 + \Tr \ThetaRP +  O(\hdag^2) \ ,
\end{align}
or $D(\vect X, \vect \Pi) = 1 - \Tr\ThetaRP + O(\hdag^2)$. In non-Abelian situations the non-trivial matrix structure of the Berry curvature will be inherited by $D$ which will be accounted for by performing the integration before taking the trace, thus yielding
\begin{align}
\label{eq:PhaseSpaceIntegralKineticVariables}
\avg{\calS} = \Tr_{N} \left\{  \int \frac{\upd^\dm R\ \upd^\dm P}{(2\pi\hbar)^\dm} \  D(\vect X, \vect \Pi) \ \bar \rho(\vect X, \vect \Pi) \bar \calS(\vect X, \vect \Pi) \right\} \ .
\end{align}
$D$ describes for example charge accumulation in the case of a topological insulator with a magnetization structure induced by ferromagnetic exchange. This effect of the Berry curvature $D$ on the density of states has been already discovered by Xiao and coworkers \cite{Xiao2005}.

Let us look at this in another way by using result (\ref{eq:BackTransformationRelation}) to transform an observable $\bar\calO$ back into the original frame. In addition, we let $\calS$ be a general observable that can posses an arbitrary matrix structure, so that contrary to relation (\ref{eq:TransformationRelationKinetic}), the additional dipole term becomes relevant.
In the end, we want to establish the connection with (\ref{eq:PhaseSpaceIntegralKineticVariables}), so we are interested in expectation values or phase-space densities (which then are quasi-probability distributions as discussed later),
\begin{align}
s(\vect x, \vect \pi) &= \Tr_N \left( \frac12 \left\{ \rho \stackrel{\Moyal}, \calS \right\} \right) = \Tr_N \calO  \ .
\end{align}
We identify $\bar\calO = \frac12\left\{ \bar\rho \stackrel{\Moyal}, \bar\calS \right\}$ and plug in the back transformation (\ref{eq:BackTransformationRelation}), so that
\begin{multline}
s(\vect x, \vect \pi) = \Tr_{N} \left\{\bar \calO + \hbar \calA_\pi \partial_{\vect x} \bar\calO - \hbar \calA_{\vect x} \partial_{\vect \pi} \bar \calO \right\} \nonumber\\
+i\hbar \Tr \left\{ \bar \calO \left[ \calA_{\vect x}, \calA_{\vect \pi} \right] \right\} + O(\hdag^2) \ .
\end{multline}
According to the discussion above, it is reasonable to assume that our observable is given as a function of the kinetic variables, i.e. $\bar\calO(\vect X, \vect \Pi)$
and it is instructive to treat the inter-band and the intra-band contributions separately by splitting $\bar\calO = \calP_{\rm d} \bar\calO \calP_{\rm d} + \bar\calO^{\rm (o)} \equiv \bar\calO^{\rm (d)} + \bar\calO^{\rm (o)}$ and likewise for $\bar\calS = \bar\calS^{\rm (d)} + \bar\calS^{\rm (o)}$, so that the contribution from the diagonal part becomes\footnote{Since $\calO^{\rm (d)}$ effectively projects the Berry connection matrices $\Ad{}$, so that we have $[ {\mathcal {\bar S}}^{\rm (d)}, \Ad{}] = 0$, the Taylor expansion is mathematically well-defined. In the case of Abelian gauge fields, the situation is trivial, while in the non-Abelian case, we need to make the assumption not to probe the internal, degenerate structure and accordingly, ${\mathcal {\bar S}}_0$ should be diagonal in the corresponding sub-space in order to describe an observable of this kind}
\begin{multline}
s^{\rm (d)}(\vect x, \vect \pi) =\\
\Tr_{N} \left\{ D(\vect x, \vect \pi) \ \bar\calO^{\rm (d)}(\vect X +\hbar \Ad{\vect \pi}, \vect \Pi -  \hbar \Ad{\vect x}) \right\} \\
+ O(\hdag^2) \ ,
\end{multline}
which basically undoes the variable transformation so that we go back to the canonical pair of variables and can write
\begin{align}
\label{eq:TrInOrigFrameDiag}
s^{\rm (d)}(\vect x, \vect \pi) = \Tr_{N} \! \left\{ \bar\rho(\vect x, \vect \pi) \  D(\vect x, \vect \pi) \ \bar\calS^{\rm (d)}(\vect x, \vect \pi) \right\} + O(\hdag^2) \,.
\end{align}
In addition, we rewrote the last term with the help of identity (\ref{eq:BerryConnectionRelation}), $i \left[ \calA_{\vect x}, \calA_{\vect \pi} \right] = \partial_{\vect x} \calA_{\vect \pi} - \partial_{\vect \pi} \calA_{\vect x} $ and, according to our previous discussion, we have $\Theta^{\scriptscriptstyle \epsilon t} = 0$, as $\calU_0$ was assumed to not explicitly depend on the energy parameter so we can replace this term with $\Tr \ThetaRP$. This contribution gives rise to the correction factor $D(\vect x, \vect \pi)$ that we already encountered before, and thus, the last result is consistent with relation (\ref{eq:PhaseSpaceIntegralKineticVariables}).

The implications of the diagonal part of $\bar\calS$ can be summarized as undoing the parameter transformation together with the appearance of the correction factor $D(\vect x, \vect p)$ which locally changes the density. However, it is not always possible to ignore the off-diagonal part of the observable $\bar\calS$, one prominent example will be the current density (see Eq. \ref{eq:CurrentOffdiagonalPart}). With a series of straightforward manipulations involving the cyclic property of the trace along with identity (\ref{eq:BerryConnectionRelation}), we eventually arrive at
\begin{multline}
\label{eq:TrInOrigFrameOffDiag}
s^{\rm (o)}(\vect x, \vect \pi) = \partial_{\vect x} \Tr_N \bar\rho \frac\hbar2 \!\left\{ \calA_{\vect \pi}, \bar\calS^{\rm (o)} \right\} \\
 - \partial_{\vect \pi} \Tr_N \bar\rho \frac\hbar2 \!\left\{ \calA_{\vect x}, \bar\calS^{\rm (o)} \right\} +O(\hdag^2) \ ,
\end{multline}
so that both contributions to the expectation value, (\ref{eq:TrInOrigFrameDiag}) and (\ref{eq:TrInOrigFrameOffDiag}) together read
\begin{multline}
\label{eq:ExpectationvalueInRotFrame}
s = \Tr_{N} \bar\rho \, D \bar\calS^{\rm (d)} + \partial_{\vect x} \Tr_N \bar\rho \, \frac\hbar2 \!\left\{ \calA_{\vect \pi}, \bar\calS^{\rm (o)} \right\} \\
 - \partial_{\vect \pi} \Tr_N \bar\rho \, \frac\hbar2 \!\left\{ \calA_{\vect x}, \bar\calS^{\rm (o)} \right\} + O(\hdag^2)\ .
\end{multline}
The importance of these last two terms will become clearer in section \ref{sec.currentDefinition} when discussing the kinetic equations of the effective theory.

To summarize this section, we have seen that when we use the kinetic terms $\vect X$ and $\vect \Pi$ as basic quantities for our observables, we end up with expressions that are manifest gauge invariant (c.f. Eqns (\ref{eq:MoyalTransform}) and (\ref{eq:PhaseSpaceIntegralKineticVariables})).
In fact, these kinetic variables appear consistently in virtually all equations of physical relevance, and moreover, it is exactly these quantities that we obtain, if we transform the canonical variables into the rotated frame,
\begin{align*}
\vect X &= \calU \Moyal \vect x \Moyal \calU^\dagger = \vect x - i\hbar \calU \Moyal \partial_{\vect \pi} \calU^\dagger \\
\vect \Pi &= \calU \Moyal \vect \pi \Moyal \calU^\dagger =  \vect \pi + i\hbar \calU \Moyal \partial_{\vect x} \calU^\dagger \ .
\end{align*}

\subsection{Is the diagonalization transformation canonical?}
Before continuing, we would like to point out that without the projection, $\vect X$ and $\vect \Pi$ still obey the canonical commutation relations, which can be easily seen by noting that the set of unprojected Berry curvatures vanishes according to identity (\ref{eq:BerryConnectionRelation}) (for example, (\ref{eq:EffectiveFieldBP}) or (\ref{eq:EffectiveFieldEP}) with $\calA^{\rm (d)}$ replaced by the $N \times N$ Berry connection $\calA$). This is actually not surprising, since then our unitary matrix $\calU_0$ is connected by $SU(N)$ gauge invariance to the identity transformation which clearly has vanishing Berry curvature. Or in other words, $\calU_0$ itself is a ${\rm SU}(N)$ gauge-transformation which naturally keeps the full ${\rm SU}(N)$ Berry curvature invariant.
This implies that our unitary transformation is a canonical one, however, since $\vect X$ and $\vect \Pi$ now posses a complicated matrix structure in the $N$-dimensional band space, they no longer commute with non-trivial matrices within this band space. For example, in a simple particle-hole symmetric two-band model
\begin{align}
\calH = \vect E(\vect p) \cdot \sig + V(\vect r) \mathbbm 1_2 \ ,
\end{align}
our band-diagonalized Hamiltonian has the form
\begin{align}
\bar \calH = E(\vect p) \sigma_z + V(\vect R) \mathbbm 1_2 \ ,
\end{align}
while $\vect R = \vect r + \hbar \calA_{\vect p}$ acquires a $2\times2$ Berry connection matrix, so that $\left[\vect R_i, \vect R_j\right] = 0 = \left[\vect P_i, \vect P_j\right]$ and $\left[\vect R_i, \vect P_j\right] = i\hbar \delta_{ij}$. Instead, the commutator $\left[ \vect R, \sigma_z \right] = \hbar \left[ \vect \calA_{\vect p}, \sigma_z \right] \neq 0$ now encodes the complicated dynamics of inter-band scattering, making the problem as a whole not easier tractable so, in the general case, the only way out is the truncation scheme. And only due to the restriction of the Berry connection matrices into a certain sub-space do the corresponding Berry curvatures yield a non-vanishing value.

\subsection{Electronic Spectrum and Magnetic Dipole Energy}

In course of the preceding discussion, we have seen that rewriting the Hamiltonian in terms of kinetic variables would render it gauge invariant. However, in order to calculate the electronic spectrum, one essentially has to fully diagonalize it, and which has to be done in terms of canonical variables. But it turns out that this remaining gauge-dependence of the Hamiltonian in the canonical representation would only affect the wavefunctions or quantities that build upon them like the retarded Greens function or the density operator. These objects will acquire local phase-factors that depend on $\vect x$ and $\vect p$ and we will illustrate this point later by explicitly studying the situation in the case of the Dirac equation. At this time, we can conclude that the electronic spectrum of the system is also gauge-independent.

Let us briefly discuss the last term of the transformation equations like (\ref{eq:effectiveHamiltonian}), and that we ignored by now,
\begin{align}
\bar\calH_{\calM} = \frac{i\hbar}4 \calP_{\rm d} \left(\left\{\calA_{\vect r}, \left[\calA_{\vect p}, \bar\calH_0\right] - \left\{\calA_{\vect p}, \left[\calA_{\vect r}, \bar\calH_0\right]\right\} \right\}  \right) \calP_{\rm d} \ ,
\end{align}
which gives rise to corrections due to virtual transitions to other bands. The term "magnetic" comes from the fact that $\bar\calH_{\calM}$ corresponds to the energy of a magnetic dipole in an external magnetic field, where the magnetic dipole is an intrinsic property of the band. For example, in the case of the Dirac equation, this term becomes essentially the magnetic Zeeman term, though in other scenarios, one obtains generalizations thereof, and even in the absence of external magnetic fields, $\bar\calH_{\calM}$ can be non-zero.

Since we will encounter terms like in $\bar\calH_{\calM}$ later, let us define the quantities ($\alpha,\beta$ are any combinations of $(t, \vect r, \vect p)$)
\begin{align}
\label{eq:DefinitionOmega}
\myTens \Omega_{\alpha,\beta} = \frac{i\hbar}2 \calP_{\rm d} \left\{ \calA_\alpha , \left[ \calA_\beta, \bar\calH_0 \right] \right\} \calP_{\rm d} \ ,
\end{align}
which is an antisymmetric tensor in $(t,\vect r,\vect p)$-space,
and in addition, it is a band-diagonal matrix, or block diagonal according to the structure defined by the projector $\calP_{\rm d}$.
The explicit structure of $\myTens \Omega$ is
\begin{align}
\label{eq:OmegaTensorDefinition}
\myTens \Omega = \begin{pmatrix} 0 & -\vect \OTR &  -\vect \OTP \\ \vect \OTR & \epsilon_{ijk} \ORR_k & \OmegaRP  \\ \vect \OTP &  -(\OmegaRP)^T &  \epsilon_{ijk} \OPP_k \end{pmatrix} \ ,
\end{align}
which has been defined in analogy to the Berry curvatures $\myTens \Theta$ given in (\ref{eq:ThetaTensorDefinitionMatrixForm}).

Now, we can express the energy term as
\begin{align}
\label{eq:HLinkOmega}
\bar\calH_{\calM} = \Tr \OmegaRP \ ,
\end{align}
where the trace is only with respect to coordinates, and not band indices, and we write explicitly
\begin{align}
\label{eq:DefinitionOmegaRP}
\OmegaRP_{ij} = \frac{i\hbar}2 \calP_{\rm d} \left\{ \calA_{r_i} , \left[ \calA_{p_j}, \bar\calH_0 \right] \right\} \calP_{\rm d} \ .
\end{align}
As will become more apparent later, terms involving the $\myTens \Omega$ tensor are related to circular currents and give rise to important terms that should not be ignored.

For the sake of completeness, and in order to establish the link to other treatments in literature (for example Ref. \onlinecite{RevModPhysXiao2010} and references therein), let us express our quantities $\myTens \Theta$ and $\myTens \Omega$ in terms of Bloch functions $(\calA_{\vect p})_{ij} = \bra{u_i} i\partial_{\vect p} \ket{u_j}$, with the Bloch band indices $i$ and $j$.
We start by introducing the gauge-invariant transition elements
\begin{align}
\Gamma_{ij}^{(\alpha,\beta)} & \equiv -2\hbar\Im\left\{ (\calA_{\alpha})_{ij} (\calA_{\beta})_{ji} \right\} \nonumber\\
&= -2\hbar\Im\left\{ \bra{u_i} i\partial_\alpha \ket{u_j} \ \bra{u_j} i\partial_\beta \ket{u_i} \right\} \ ,
\end{align}
which are anti-Hermitian, $\Gamma_{ij}^{(\alpha,\beta)} = -\Gamma_{ji}^{(\alpha,\beta)}$ due to the Hermiticity of $\calA$ and, as the name suggests, they describe corrections due to virtual transitions between band $i$ and $j$.
Then the Berry curvatures projected onto band $i$ can be expressed as
\begin{align}
\label{eq:DefinitionFullTheta}
\myTens \Theta_{\alpha\beta} = \hbar (\partial_\alpha \Ad\beta - \partial_\beta \Ad\alpha)_{ii} = \sum_j \Gamma_{ij}^{(\alpha,\beta)} \ ,
\end{align}
and one can interpret the Berry connection as being corrections to the kinetic variables $\vect x, \vect \pi$ due to virtual transitions into all other bands $j$. The sum rule (\ref{eq:SumRule}) is then a direct consequence of the anti-Hermiticity of $\Gamma$.
In the same spirit, the dipole terms projected on Bloch band $i$ read
\begin{align}
\myTens \Omega_{\alpha\beta} = \frac12 \sum_j \left( \epsilon_i - \epsilon_j \right) \Gamma_{ij}^{(\alpha,\beta)} \ ,
\end{align}
where the energies are $\epsilon_i \equiv (\bar\calH_0)_{ii}$.

As might be apparent from the above expressions, there exists a direct link between $\myTens \Theta$ and $\myTens \Omega$ in the case of a two-level system, and which we establish in the following. First, a general diagonal 2-level Hamiltonian $\bar\calH_0$ can be divided into a symmetric part $\propto \mathbbm 1_2$ which consequently commutates out in expression (\ref{eq:DefinitionOmega}), and an antisymmetric part $\propto \sigma_z$ responsible for a finite contribution. After some simple algebra, we find $\Tr_2 \sigma_z \myTens \Omega = 0$ which implies that $\myTens\Omega$ is band diagonal, and the two bands get the same contribution, in particular the magnetic dipole energy is the same for the two bands. The diagonal part is obtained by taking the trace and using its cyclic properties,
\begin{align}
\Tr_2 \myTens \Omega_{\alpha,\beta} &= i\hbar \Tr_2 \bar\calH_0 \left[ \calA_{\alpha}, \calA_{\beta} \right] \nonumber\\
&= \hbar \Tr_2 \bar\calH_0 (\partial_{\alpha} \Ad{\beta} - \partial_{\beta}  \Ad{\alpha} )
=\Tr_2 \left( \bar\calH_0 \myTens \Theta_{\alpha,\beta} \right) \ ,
\end{align}
where in the last two steps, we used the identity (\ref{eq:BerryConnectionRelation}) along with definition (\ref{eq:DefinitionFullTheta}) to replace the commutator.
We can finally cast the result into the form
\begin{align}
\label{eq:OmegaThetaTwoLevel}
\myTens \Omega = \frac12 \Tr_2 \left( \bar\calH_0 \myTens\Theta \right) \ \mathbbm 1_2 \ ,
\end{align}
where in this expression, the diagonal contribution of $\bar \calH_0$ drops out because of the sum rule $\Tr_2 \myTens\Theta = 0$.

In particular, we can express the magnetic dipole energy as
\begin{align}
\label{eq:MagnetizationEnergyTwoLevel}
\bar\calH_{\calM} = \frac12 \Tr_2 \left( \bar\calH_0 \Tr\ThetaRP \right) \ \mathbbm 1_2 \ ,
\end{align}
where $\Tr\ThetaRP \equiv \sum_i \ThetaRP_{ii}$, whereas $\Tr_2$ denotes the trace with respect to the two level band space.

\section{Equations of Motion}
\label{sec.EffectiveTheoryEquationsOfMotion}
We are now interested in the dynamics of the systems, especially in a consistent effective descriptions of the physics within a certain band.

\subsection{Quasi-probability distributions and currents}
\label{sec.currentDefinition}

Within the Wigner framework, the basic quantity is the quasi-probability distribution $\rho(\vect r, \vect p)$, while the usual momentum and position probability distributions can be generally defined as marginals
\begin{align*}
\rho(\vect r) = \int \frac{\upd^\dm p}{(2\pi \hbar)^\dm}\  \rho(\vect p, \vect r) \ , \\
\rho(\vect p) = \int \frac{\upd^\dm r}{(2\pi \hbar)^\dm} \ \rho(\vect p, \vect r) \ ,
\end{align*}
and they are linked to the probability interpretation of quantum mechanics, i.e. $\rho(\vect r) = \abs{\psi(\vect r)}^2$ and $\rho(\vect p) = \abs{\psi(\vect p)}^2$. In the rotated frame however, one has to express them in terms of kinetic variables instead, in order to obtain gauge invariant results.

Now the goal which we want pursue in the following is to find proper quantities in the rotated frame that lead to a consistent physical description when projected onto a certain band.

Since the velocity operator is given by $$\frac{\upd \vect R}{\upd t} = \frac1{i\hbar} \left[ \bar\Xi \stackrel{\MoyalKinetic}, \vect R \right] \ ,$$ and inspired by result (\ref{eq:ExpectationvalueInRotFrame}) for the expectation values in the rotated frame, we define the quasi-probability density $\mathfrak n$, current density $\mathfrak {\vect j}$ and force density $\mathfrak {\vect q}$ as follows
\begin{widetext}
\begin{align}
\label{eq:QuasiProbDensity}
\mathfrak n(\vect R, \vect P, t) &\equiv \Tr_N \left\{ \bar\rho \ D \right\} \ , \\
\label{eq:QuasiProbCurrentR}
\mathfrak {\vect j}(\vect R, \vect P, t) &\equiv \Tr_N \left\{\bar\rho \  D \frac{\upd \vect R}{\upd t}  \right\} - \vect \nabla_{\!\vect R} \xproduct \Tr_N \bar\rho \vect \OPP - \vect\nabla_{\! \vect P} \sproduct \Tr_N \bar\rho \OmegaRP \ , \\
\label{eq:QuasiProbCurrentP}
\mathfrak {\vect q}(\vect R, \vect P, t) &\equiv \Tr_N \left\{\bar\rho \ D \frac{\upd \vect P}{\upd t}  \right\} - \vect \nabla_{\!\vect P} \xproduct \Tr_N \bar\rho \vect \ORR + \vect\nabla_{\! \vect R} \sproduct \Tr_N \bar\rho (\OmegaRP)^T  \ ,
\end{align}
\end{widetext}
where we used that $\OmegaPR =- (\OmegaRP)^T$.
These densities obey a conservation law in phase space, or Liouville's theorem that states
\begin{align}
\label{eq:LiouvilleTheorem}
\partial_t \mathfrak n(\vect R, \vect P, t) + \vect \nabla_{\! \vect R} \ \mathfrak {\vect j}(\vect R, \vect P, t) + \vect \nabla_{\! \vect P} \ \mathfrak {\vect q}(\vect R, \vect P, t) = 0 \ .
\end{align}
We substitute the definitions (\ref{eq:QuasiProbDensity})-(\ref{eq:QuasiProbCurrentP}) into Liouville's theorem and establish the identity (\ref{eq:LiouvilleTheorem}) after some algebra, using the kinetic equation (\ref{eq:QuantumBoltzmannEquation}) for $\bar\rho$ and the equality of second partial derivatives and dropping terms of order $O(\hbar^2)$. Furthermore, we need to use the following identities between Berry curvatures
\begin{align}
\label{eq:IdentitiesBetweenBerryCurvatures}
\partial_t D - \vect \nabla_{\! R} \vect \EP + \vect \nabla_{\! P} \vect \ER &= 0 \ , \nonumber\\
\partial_{R_i} D + \partial_{R_k} \ThetaRP_{ik} - (\vect \nabla_{\! P} \xproduct \vect \BR)_i &= 0 \ , \nonumber\\
- \partial_{P_i} D - \partial_{P_k} \ThetaRP_{ki} - (\vect \nabla_{\! R} \xproduct \vect \BP)_i &= 0 \ ,
\end{align}
which can be readily shown by plugging in the definitions (\ref{eq:EffectiveFieldBR})-(\ref{eq:DefinitionThetaRP}), (\ref{eq:EffectiveFieldER}) and (\ref{eq:EffectiveFieldEP}).
This result strongly emphasizes the importance of including the correction factor $D(\vect R, \vect P, t)$ into the expectation values.

Here, it is important to realize that the current matrices have an off-diagonal structure that one has to take into account in the light of expression (\ref{eq:ExpectationvalueInRotFrame}) and which eventually leads to the last two terms in the current densities (\ref{eq:QuasiProbCurrentR}) and (\ref{eq:QuasiProbCurrentP}), and constitute divergence-free, or circular currents. Explicitly for the 4-component current, this off-diagonal part takes the form
\begin{align}
\label{eq:CurrentOffdiagonalPart}
\left( \frac{\upd \vect X}{\upd t} \right)_{\rm o} = -i \left[ \bar \calH_0 , \calA_{\vect \pi} \right]
\end{align}
and is completely expressed in terms of quantities of the rotated frame and arises due to the off-diagonal elements of the Berry connection matrix.

Just like the Wigner function is a quasi-probability and can be given physical sense only after taking expectation values, the same applies to $\mathfrak n$, $\mathfrak {\vect j}$ and $\mathfrak {\vect q}$. In particular, physical meaning can be given only to quantities like $n(\vect R)$ or $\vect j(\vect R)$. This is related to Heisenberg's uncertainty which states that momentum and position uncertainty have to be larger than Plank's constant $\hbar$, i.e. $\Delta P \Delta R \gtrsim \hbar$. The same is true for the conjugate variables time and energy.

Liouville's theorem constitutes a conservation law for the quasi-probability densities, and one can for example integrate (\ref{eq:LiouvilleTheorem}) over all momenta in order to find a continuity equation for the probability densities of the particle and current densities,
\begin{align}
\partial_t n(\vect R, t) + \vect \nabla_{\!R} \vect j(\vect R, t) = -\oint_{\partial V} \upd^{\dm-1} \vect P \  \mathfrak {\vect q}(\vect R, \vect P, t) \ ,
\end{align}
where the closed surface integral on the right-hand side has been obtained by virtue of Gauss's theorem, and it vanishes if we assume that there is no net momentum current flow through the surface at infinity. Explicitly, the particle density is
\begin{align}
\label{eq:ParticleDensityDiagFrame}
n(\vect R, t) = \int \frac{\upd^\dm P}{(2\pi \hbar)^\dm} \  \Tr_N \left(\bar\rho \ D \right) \ ,
\end{align}
and likewise for the current density,
\begin{widetext}
\begin{align}
\label{eq:CurrentDensityDiagFrame}
\vect j(\vect R, t) = \int \frac{\upd^d P}{(2\pi \hbar)^d} \  \Tr_N \left(\bar\rho \  D \frac{\upd \vect R}{\upd t}  \right) - \vect \nabla_{\!\vect R} \xproduct  \int \frac{\upd^d P}{(2\pi \hbar)^d} \  \Tr_N \bar\rho \vect \OPP - \oint \frac{\upd^{\dm-1} \vect S_P}{(2\pi\hbar)^d} \Tr_N \bar\rho \OmegaRP \ ,
\end{align}
\end{widetext}
while the previous discussion has shown that these definitions are in fact meaningful physical quantities and constitute the central result of this section. We will see interesting implications of the additional dipole term in the current when studying the examples of the Dirac equation.

For the sake of completeness, we use result (\ref{eq:ExpectationvalueInRotFrame}) to define densities corresponding to the physical quantity $\calS$ in the diagonalized frame
\begin{widetext}
\begin{align}
\label{eq:ObservableDiagFrame}
S(\vect R, t) \equiv \int \frac{\upd^\dm P}{(2\pi\hbar)^\dm} \  \Tr_{N} \left(\bar\rho \  D \bar\calS - \vect \nabla_{\! \vect R} \ \bar\rho \frac\hbar2\! \left\{ \calA_{\vect P} , \bar\calS^{\rm (o)} \right\} \right) +
\oint \frac{\upd^{\dm-1} \vect P}{(2\pi\hbar)^\dm} \Tr_N \bar \rho \frac\hbar2\! \left\{ \calA_{\vect R} , \bar\calS^{\rm (o)} \right\}  \ ,
\end{align}
\end{widetext}
and likewise for $S(\vect P)$. This expression is consistent with the definition of a density operator $\delta(\vect r -\vect r_0) S(\vect r, \vect p)$ which, when transformed into  the rotated frame yields the given result. Furthermore, all the given expressions can still explicitly depend on energy via the gauge invariant parameter $E$, which however does not affect the discussion here.

\subsection{Operator equations of motion}
In order to calculate the densities (\ref{eq:QuasiProbCurrentR}) and (\ref{eq:QuasiProbCurrentP}), we need to evaluate the equations of motion for kinetic position $\vect R$ and momentum $\vect P$.
Using this result and (\ref{eq:MoyalTransform}), we can now immediately write down the equations of motion
\begin{align}
\label{eq:SemiclassicalEquationsOfMotionR1}
\frac{\upd \vect R}{\upd t} &= \frac1{i\hbar} \left[\bar \Xi \stackrel{\Moyal}{,} \vect R \right] = \frac{\partial \bar \calH} {\partial \vect P} \left(\myTens 1 + \ThetaRP\right) - \vect \EP +  \frac{\partial \bar \calH} {\partial \vect R} \xproduct \vect \BP \\
\label{eq:SemiclassicalEquationsOfMotionP1}
\frac{\upd \vect P}{\upd t} &= \frac1{i\hbar} \left[\bar \Xi \stackrel{\Moyal}{,} \vect P \right] = -\left(\myTens 1 + \ThetaRP\right) \! \frac{\partial \bar \calH} {\partial \vect R} + \vect \ER \! +  \frac{\partial \bar \calH} {\partial \vect P} \xproduct \vect \BR \, ,
\end{align}
or, to leading order in $\hdag$,
\begin{align}
\label{eq:SemiclassicalEquationsOfMotionR}
\frac{\upd \vect R}{\upd t} (\myTens 1 - \ThetaRP) &= \frac{\partial \bar \calH} {\partial \vect P} - \vect \EP -  \frac{\upd \vect P}{\upd t} \xproduct \vect \BP \\
\label{eq:SemiclassicalEquationsOfMotionP}
(\myTens 1 - \ThetaRP) \frac{\upd \vect P}{\upd t} &= -\frac{\partial \bar \calH} {\partial \vect R} + \vect \ER +  \frac{\upd \vect R}{\upd t} \xproduct \vect \BR \ ,
\end{align}
where we see that the effect of $\ThetaRP$ is related to a change of phase-space in the course of the diagonalization transformation. The beauty of this result is the symmetry in which effective magnetic and electric fields appear in these equations.

We are now performing a quantum average of the kinetic equations (\ref{eq:SemiclassicalEquationsOfMotionR}) and (\ref{eq:SemiclassicalEquationsOfMotionP}) with respect to a density matrix that is peaked around a certain value $\vect P_c$ and $\vect R_c$, which in effect corresponds to a Gaussian wave-packet that is well beyond the limits of the Heisenberg uncertainty. Then our equations of motion (\ref{eq:SemiclassicalEquationsOfMotionR}) and (\ref{eq:SemiclassicalEquationsOfMotionP}) essentially become classical ones with all kinetic variables replaced by $P_c$ and $R_c$. These equations of motion for the center of mass coordinates of a wave packet have been directly obtained by Sundaram and coworkers \cite{Sundaram1999}.
The circular current terms in (\ref{eq:QuasiProbCurrentR}) and (\ref{eq:QuasiProbCurrentP}) vanish for the wave packet, which is to be expected physically, since this description reduces the electron to a point particle with coordinates $\vect P_c$ and $\vect R_c$, and which does not posses any internal motion described by the circular currents. Mathematically, after integration over the whole phase-space, one can rewrite them in terms of surface integrals which vanish for the well-localized wave-packet.

\subsection{Non-equilibrium description}
Now, it is just a matter of finding the density matrix $\bar\rho(\vect R, \vect P, t)$ in order to explicitly calculate anything within this framework, in particular, when one is interested in non-equilibrium phenomena. One way to proceed is within Keldysh formalism and to consider the kinetic equation for the lesser Greens function \cite{Rammer1986},
\begin{align}
\left[ \Xi \stackrel{\Moyal}{,} \calG^{<} \right] = 0 \ ,
\end{align}
which, after transformation into the rotated frame, can be evaluated to leading order correction in $\hdag$ by using (\ref{eq:MoyalTransform}) which by virtue of (\ref{eq:SemiclassicalEquationsOfMotionR1}) and (\ref{eq:SemiclassicalEquationsOfMotionP1}) can be recast into
\begin{widetext}
\begin{align}
\label{eq:QuantumBoltzmannEquation}
\left[ \partial_t  + \frac{\upd \vect R}{\upd t} \vect \nabla_{\! R} + \frac{\upd \vect P}{\upd t} \vect \nabla_{\! P} + \left(\frac{\partial \bar\calH}{\partial t} + \vect \ER \frac{\partial \bar\calH}{\partial \vect P} - \vect \EP \frac{\partial \bar\calH}{\partial \vect R} \right) \partial_E \right] \bar\calG^{<} = 0 \ .
\end{align}
\end{widetext}

When we integrate the lesser Greens function over energy, we obtain the density matrix
\begin{align}
\bar\rho(\vect R, \vect P, t) = \int \frac{\upd E}{2\pi} \ \bar\calG^{<}(\vect X, \vect \Pi) \ ,
\end{align}
and, since the terms proportional to $\partial_E \bar\calG^{<}$ vanish after integrating (\ref{eq:QuantumBoltzmannEquation}) over all energies, we are left with the following differential equation for the density matrix,
\begin{align}
\label{eq:BoltzmannEquation}
\left( \partial_t  + \frac{\upd \vect R}{\upd t} \vect \nabla_{\! R} + \frac{\upd \vect P}{\upd t} \vect \nabla_{\! P} \right) \bar\rho(\vect R, \vect P,t) = 0 \ ,
\end{align}
which is nothing but a Boltzmann equation including quantum corrections in terms of Berry curvatures.
In the end, it is quite analogous to the quantum Boltzmann equation along with the gradient expansion we used Ref. \onlinecite{Wickles2009}. Of course, the major difference is the absence of the collision integral in the present formulation, however it is possible to include it here as well, which is however left for future investigations. Of course, $\bar\rho$ can additionally include discrete quantum degrees of freedom like the spin, which means that our rotated frame is block-diagonal and so is $\bar\rho$.

We now show how to obtain the distribution function, when it is known at some initial time, $\bar\rho(\vect R, \vect P, t=0)$. In general, these problems can be formulated in terms of an inhomogeneous Boltzmann equation with a source term $S(\vect R, \vect P, t)$,
\begin{align}
\label{eq:QuantumBoltzmannEquationInhomogeneous}
\left( \partial_t  + \frac{\upd \vect R}{\upd t} \vect \nabla_{\! R} + \frac{\upd \vect P}{\upd t} \vect \nabla_{\! P} \right) \bar\rho(\vect R, \vect P,t) = S(\vect R, \vect P, t) \ .
\end{align}
To proceed in the usual way, we define the Greens function $\calG_{\rm c}$,
\begin{multline}
\left( \partial_t  + \frac{\upd \vect R}{\upd t} \vect \nabla_{\! R} + \frac{\upd \vect P}{\upd t} \vect \nabla_{\! P} \right) \calG_{\rm c}(\vect R, \vect P, t, \vect R', \vect P', t') \\
 = \delta(t - t') \delta(\vect R - \vect R') \delta(\vect P - \vect P') \ ,
\end{multline}
for which we find the general solution
\begin{multline}
\label{eq:GeneralBallisticGreensFunction}
\calG_{\rm c}(\vect R, \vect P, t, \vect R', \vect P', t') \\
= \Theta(t-t') \ \delta(\vect R - \vect R_{\rm c}(t)) \ \delta(\vect P - \vect P_{\rm c}(t)) \ ,
\end{multline}
where $\Theta$ is the Heaviside step function, and $\vect R_{\rm c}(t)$ and $\vect P_{\rm c}(t)$ describe the classical orbits which obey the equations of motion
\begin{align*}
\vect {\dot R}_{\rm c} &= \frac{\upd \vect R}{\upd t} &\quad \vect R_{\rm c}(t') &= \vect R'  \ , \\
\vect {\dot P}_{\rm c} &= \frac{\upd \vect P}{\upd t} &\quad \vect P_{\rm c}(t') &= \vect P' \ .
\end{align*}
This solution describes ballistic trajectories given by the kinetic equations (\ref{eq:SemiclassicalEquationsOfMotionR1}) and (\ref{eq:SemiclassicalEquationsOfMotionP1}) and which including the effects of the Berry curvatures, while the particle being initially at phase space coordinates $(\vect R', \vect P', t')$. For example, in the absence of any spatially dependent potential, one simplyhas straight lines described by the Greens function $\calG_{\rm c} = \Theta(t-t_0) \  \delta(\vect R - \vect V (t-t_0)) \ \delta(\vect P - \vect P')$, where $\vect V = \frac{\partial \bar\calH_0}{\partial \vect P}$ is the group velocity. This is analogous to other quasiclassical equations like the Eilenberger equation which can be also described in terms of classical trajectories  \cite{Shelankov1985,Rammer1986}.

Now, the solution is readily given by
\begin{widetext}
\begin{align}
\bar\rho(\vect R, \vect P, t) = \int_{-\infty}^{+\infty}  \upd t' \int \upd^\dm R' \int \upd^\dm P' \ \calG_{\rm c}(\vect R, \vect P, t, \vect R', \vect P', t') \ S(\vect R', \vect P', t') \ ,
\end{align}
\end{widetext}
where, due to the delta functions in $\calG_{\rm c}$, we get contributions only from trajectories which end at $(\vect R, \vect P)$ at time $t$. Since the trajectory is well defined by the given pair $(\vect R, \vect P)$ already, we can find all corresponding points $(\vect R', \vect P')$ by going back in time so we can eventually rewrite the above integral as
\begin{align}
\label{eq:SolutionInhomogeneousBoltzmannEquation}
\bar\rho(\vect R, \vect P, t) = \int_{-\infty}^{t}  \upd t' \ S(\vect R_{\rm c}(t'), \vect P_{\rm c}(t'), t') \ ,
\end{align}
where now, the classical orbits are defined such that its position in phase space at time $t$ is $\vect R_{\rm c}(t) = \vect R$ and $\vect P_{\rm c}(t) = \vect P$.

Let us now go back to the original initial value problem which we can easily solve by using the source term $\delta(t) \bar\rho(\vect R, \vect P, t=0)$, so that the general solution reads
\begin{align}
\bar\rho(\vect R, \vect P, t) = \Theta(t) \ \bar\rho(\vect R'_{\rm c}(0), \vect P'_{\rm c}(0), 0)  \ ,
\end{align}
which has the desired properties for $t \ge 0$. This indeed corresponds to a motion in phase space as an incompressible fluid, as stated by Liouville's theorem.

\subsection{Polarization}
\label{sec.Polarization}

Another way to interpret the kinetic variables is obtained by considering the electrical polarization which we pursue according to the pioneering work of Vanderbilt and coworkers \cite{Vanderbilt1993}. They derive the polarization in terms of adiabatic transport, where some perturbation that leads to a polarization in the crystal is adiabatically turned on.
The current due to adiabatic transport is \cite{Thouless1983}
\begin{align}
\vect j = \int  \frac{\upd^\dm r \ \upd^\dm p}{(2\pi\hbar)^\dm} \ \Tr_{N} \frac{\upd \vect R}{\upd t} = - \int_{\rm BZ}  \frac{\upd^\dm p}{(2\pi\hbar)^\dm} \ \Tr_{N} \vect \EP \ ,
\end{align}
where integration is only within the Brillouin zone. The electric polarization is then obtained by integrating over time,
\begin{align}
\label{eq:DefinitionPolarization}
\vect {\mathfrak P}(\vect R) = \int \upd t \  \vect j(\vect R, t) \ .
\end{align}
The definition as an adiabatic transport process is necessary in order to obtain a truly gauge invariant result \cite{Vanderbilt1993}. In the periodic gauge however \cite{RevModPhysXiao2010}, it is nevertheless possible to construct
\begin{align}
\vect {\mathfrak P} = \int \frac{\upd^\dm p}{(2\pi\hbar)^\dm} \ \calA_{\vect p} \ ,
\end{align}
but in the general case, it is required to consider the full Berry connection structure (including in particular $\calA_t$) in order to obtain the correct result for the polarization.
In the same spirit, one could define a polarization in momentum space which can be thought of as a Doppler shift in the rotated frame or also termed anomalous velocity.

In this sense, one could loosely interpret these Berry connections as shifts that the canonical variables acquire and which depend on full phase space, albeit these shifts are not directly physically observable, only when one does an integration with respect to either the position or the momentum variable one obtains the observable electric polarization in real space or reciprocal space, respectively. This is analogous to the quasi-probability distribution $\rho(\vect r, \vect p)$ in the Wigner picture, which can be given only physical interpretation as probability density when integrated either over whole momentum space or real space. However, if we try to transfer this idea, we still will have a shift that depends on the frame, i.e. it is still not gauge invariant. Only for a very specific gauge, one obtains the physical polarization.

\subsection{Bulk-boundary correspondence}
\label{sec.BulkBoundary}
It is interesting to note that the circular current term
\begin{align}
- \vect \nabla_{\!\vect R} \xproduct  \int \frac{\upd^d P}{(2\pi \hbar)^d} \  \Tr_N \bar\rho \vect \OPP
\end{align}
appearing in the current density (\ref{eq:CurrentDensityDiagFrame}) is a manifestation of the bulk-boundary correspondence. \cite{Essin2011}
In the homogeneous bulk, these circular currents compensate each other and yield zero, but at the edge of the system, there will be residual currents flowing on the surface. Due to the circular nature of these edge currents, the associated edge states are also of topological nature. Therefore, a nontrivial (bulk) value of $\vect \OPP$ indicates a topological state of matter.

\section{Hierarchy of Effective Theories -- Inclusion of the Electromagnetic Field}
\label{sec.HierarchyAndElectromagneticField}
In many cases, one starts from an effective theory which requires dealing with gauge invariant momentum and/or position operators in the original frame. This might happen if one uses an effective theory derived from a more comprehensive theory; or more commonly, one simply wishes to include an external electromagnetic field.
Qualitatively, one expects no new concepts or phenomena to emerge in effective theories at a lower level in the hierarchy, as one could also go directly from the topmost theory to the lowest in a single step.

Here, our focus lies on the combined effect of the electromagnetic field and the Berry connections emerging from the diagonalization of a band Hamiltonian. In this case, we are starting from a theory which is gauge invariant in real space, thus formulated in terms of the minimally coupled momentum $\vect p - q \vect A(x)$.
Since all our derivations so far are only linear in $\hbar$, we treat the electromagnetic vector potential $q\vect A$ ($q$ is the charge of the particle, so we usually have $q=-e$) on a different level than $\hbar\calA$, thus formally taking $\hbar$ and $q$ as independent expansion parameters.
To begin with, we have the usual canonical pair $\vect x$ and $\vect \pi$ and the kinetic pair $\vect x$ and $\vect \pi + q \vect A(\vect x)$, where $\vect A = (-\phi, \vect A_{\vect r})$ is the usual 4-component vector potential of the electromagnetic field. In the following, we denote $\vect B(\vect r,t) = \vect \nabla_{\!\vect r} \xproduct \vect A_{\vect r}$ and $\vect E(\vect r,t) = -\partial_t \vect A_{\vect r} + \vect \nabla_{\!\vect r} \vect A_0$ (not to be confused with the energy $E$) as the external electric and magnetic fields.

The Hamiltonian is specified as $\calH(\vect x, \vect p - q \vect A_{\vect r}(\vect x))$ and is diagonalized by an appropriate unitary matrix $\calU(\vect x, \vect p - q \vect A_{\vect r}(\vect x))$, so that in the diagonalized frame we arrive at the kinetic pair of variables
\begin{align}
\label{eq:ParameterTransformationEMField}
\vect X &= \vect x - \hbar \calA_{\vect \pi}(\vect x, \vect p - q\vect A_{\vect r}) \nonumber \\
\vect \Pi &= \vect \pi + q \vect A(\vect X) + \hbar \calA_{\vect x}(\vect x, \vect p - q\vect A_{\vect r}) \ .
\end{align}
Since derivatives now also act on $\vect A_{\vect r}$, we have
\begin{align}
\label{eq:calASplitOffEM}
\calA_{x_k} &= \calA_{X_k} - q (\partial_{X_k} A_l) \calA_{P_l}
\end{align}
so that we can write more explicitly
\begin{align}
R &= r + \hbar \vect \calA_{\vect P} \\
E &= \epsilon - q \phi + \hbar \vect \calA_{\vect R} + \hbar q \vect E \sproduct \vect \calA_{\vect P} \\
\vect P &= \vect p - q \vect A_{\vect r}(\vect x) - \hbar \calA_{\vect R} - \hbar q \vect B \times \vect \calA_{\vect P} \ .
\end{align}
Note that $\calA_{\vect X}$, as well as $\calA_{\vect P}$ still carry the full gauge invariant momentum $\vect p - q\vect A_{\vect r}$ and for simplicity, all Berry connections involved are assumed to be Abelian.

The fictitious fields in the presence of an electromagnetic field can be derived by plugging these kinetic variables into the commutator relations
(\ref{eq:KineticCommutatorRelations}) and (\ref{eq:KineticCommutatorRelationsEField}) and keeping all terms up to order $O(\hbar)$ and quadratic in the fields $O(q^2)$. In order to obtain manifest gauge invariant results, we express everything in terms of kinetic variables by using $\partial_{x_j} = \partial_{X_j} - q (\partial_{X_j} A_l) \partial_{P_l}$ and $\vect E(\vect X) = \vect E(\vect x) + (\calA_{\vect P} \partial_{\vect r}) \vect E(\vect x)$.
Denoting primed quantities as the Berry curvatures describing the combined effect of external electromagnetic and the diagonalization, we can summarize the results as
\begin{widetext}
\begin{align}
\label{eq:FictitiousBRp}
\vect \BRp &= q \vect B(\vect R) + \vect \BR + q \left[\Tr(\ThetaRP) \vect B  - \vect B  \ThetaRP \right] + q^2  \left(\vect \BP \sproduct \vect B \right) \vect B \\
\vect \BPp &= \vect \BP \\
\label{eq:FictitiousERp}
\vect \ERp &= q \vect E(\vect X) + \vect \ER + q \ThetaRP \vect E + q\vect B \xproduct \vect \EP + q^2 \vect B \xproduct (\vect E \xproduct \vect \BP) \\
\label{eq:FictitiousEPp}
\vect \EPp &= \vect \EP + q \vect E \xproduct \vect \BP \\
\label{eq:FictitiousThetaRPp}
\ThetaRPp_{ij} &= \ThetaRP_{ij} + q \left[  \left( \vect \BP \sproduct \vect B \right) \delta_{ij} - \BP_i B_j \right] \ .
\end{align}
\end{widetext}
Here, Unprimed quantities $\vect \calB$, $\vect \calE$, $\myTens\Theta$ correspond to the Berry curvatures of the system in the diagonalized frame in absence of the external electromagnetic field $q\vect A$.

It is important to note that these results have to be substituted into the equations of motion (\ref{eq:SemiclassicalEquationsOfMotionR1}), (\ref{eq:SemiclassicalEquationsOfMotionP1}) and {\it not} (\ref{eq:SemiclassicalEquationsOfMotionR}), (\ref{eq:SemiclassicalEquationsOfMotionP})  since the latter has been obtained by dropping terms beyond leading order corrections in Berry curvatures, whereas our results for the fictitious fields are in fact higher order in Berry connections ($q^2 \hbar$ actually already corresponds to terms of third order in the expansion of the Moyal product).

The term $\vect E \xproduct \vect \BP$ of $\vect \EPp$ is the anomalous velocity term that is for example responsible for the quantum Hall effect, since the Berry curvature $\vect \BP$ becomes non-trivial when the system is in the Quantum Hall state. Similarly, in systems with spin-orbit interactions, this term constitutes the intrinsic contribution to the anomalous Hall effect. The reciprocal effect thereof is described by the term $q\vect B \xproduct \vect \EP$, where a external magnetic field transforms a momentum space electric field into a real-space one.

$\ThetaRPp$ gets modified by a magnetic field term that resembles a dipole interaction between a real-space and the momentum space magnetic field. In fact, as we have seen in result (\ref{eq:MagnetizationEnergyTwoLevel}), and as we will see explicitly at the end of this section, $\ThetaRPp$ is directly related to the magnetic dipole energy in the case of a two-level model, and thus, it is the magnetic field term in $\ThetaRPp$ which will give rise to the Zeeman energy in the Hamiltonian. Furthermore, the term quadratic in the fields can be absorbed by $\ThetaRPp$, i.e. fictitious electric field can be recast into
\begin{align}
\vect \ERp &= q \vect E + \vect \ER + q \ThetaRPp \vect E +  q\vect B \xproduct \vect \EP \ .
\end{align}
From this expression, we see that the external electric field enters effectively as $q(\mathbbm 1 + \ThetaRPp) \vect E$, and the electric field acting on the system is thus renormalized. An analogous renormalization of the external magnetic field appears also in the result for $\vect \BRp$.

The modification of the transformation rule (\ref{eq:TransformationRelation}) for arbitrary observables is easily found by virtue of (\ref{eq:calASplitOffEM}),
\begin{widetext}
\begin{align}
\bar\calO' = \bar \calO + \left[\calY_1' - \calY_1, \bar\calO_0 \right] + \frac\hbar2 q \vect E \left\{\vect \calA_{\vect P} , \partial_E \bar\calO_0 \right\} - \epsilon_{ijk} q B_k \frac{\hbar}{2} \left\{ \calA_{\vect P_i} , \left[\partial_{\vect P_j} - \tfrac i2 \calA_{\vect P_j}, \bar\calO_0\right] \right\} \ ,
\end{align}
\end{widetext}
where again, primed symbols represent quantities in presence of an external magnetic field.
In particular, evaluating the above expression for the Hamiltonian, we find for the magnetization energy in the rotated frame $\bar\calH_{\calM} = \Tr \OmegaRPp$ (see Eq. (\ref{eq:HLinkOmega})),
where
\begin{align}
\label{eq:OmegaRPp}
\OmegaRPp_{ij} = \OmegaRP_{ij} + q \left[  \left( \vect \OPP \sproduct \vect B \right) \delta_{ij} - \OPP_i B_j \right] \ .
\end{align}
Notice the resemblance between this expression and (\ref{eq:FictitiousThetaRPp}); in fact, it turns out we obtain $\myTens\Omega'$ by substituting the Berry curvatures with the analogue quantities of $\myTens\Omega$ in results (\ref{eq:FictitiousBRp})-(\ref{eq:FictitiousThetaRPp}).
In two dimensions, $\ThetaRPp_{33}$ is zero,\footnote{However, even in lower dimensional system we are still using all three dimensions in our vector calculus and according to (\ref{eq:FictitiousThetaRPp}), $\ThetaRPp_{33} \neq 0$, so we have to pay attention to these subtleties for dimensions $\le 2$. }
and thus $\Tr\ThetaRPp = \Tr\ThetaRP + q \vect B \sproduct \vect \BP$, i.e. a factor 2 less than in the 3D case, and likewise, in 1D $\Tr\ThetaRPp = \Tr\ThetaRP$.
Nevertheless, we see that the magnetic field couples in the shape of a dipole term $\vect B \sproduct \vect \BP$ to the momentum space magnetic field $\vect \BP$. Since the magnetic dipole energy is already of order $\hdag$, we can equally write $\vect B(\vect X)$ and $\vect B(\vect x)$, since substituting canonical and kinetic variables affects only higher orders in $\hdag$.

\section{The Dirac Equation - An alternative perspective on deriving the Pauli-Hamiltonian}
\label{sec.DiracEquation}
The goal of this section is illustrate our formalism, while at the same time serving as a concise summary of the framework, essentially making use of all the central results. The Dirac equation and its descendants like the Pauli Hamiltonian are the most fundamental equations in condensed matter physics, nevertheless, we present some interesting insight into the physics revealed by this rather simple equation.

We consider the Dirac Hamiltonian in the presence of a scalar potential and minimally coupled to the magnetic field \cite{SchwablQM2008},
\begin{align}
H = c \vect \alpha (\vect p - q \vect A) + m c^2 \beta + V(\vect x) \mathbbm1_4 \ ,
\end{align}
where $m$ is the rest mass and $\vect p$ the momentum of the electron. As
\begin{align}
\vect \alpha = \begin{pmatrix} 0 & \sig \\ \sig & 0\end{pmatrix} \ , \qquad \beta  = \begin{pmatrix} \mathbbm 1_2 & 0 \\ 0 & -\mathbbm 1_2 \end{pmatrix} \ ,
\end{align}
$H$ acts on the 4-dimensional Dirac Spinor and thus can be considered as a 4-band model.

The Foldy-Wouthuysen transformation \cite{FoldyWouthuysen1950} which brings the Dirac equation into diagonal form is described by the unitary matrix
\begin{align}
\calU_0 = \frac {(E_P+mc^2)\mathbbm1_4 + c  \beta \vect \alpha \vect P} {\sqrt{E_P(E_P+m c^2)}} \ ,
\end{align}
where $E_P = c \sqrt{P^2 + m^2 c^2}$ is the relativistic energy of the electron with gauge invariant momentum $P$. Performing, as outlined before (cf. Eq. (\ref{eq:effectiveHamiltonian})), the diagonalization with respect to $\calU_0$, we arrive at
\begin{multline}
\label{eq:DiracPauliHamiltonian}
\bar \calH = \calU \Moyal H \Moyal \calU^\dagger = E_P \sigma_0 \tau_z \\
  + \left( V(\vect r) \sigma_0 + \hbar (\partial_{\vect r} V) \calA_{\vect P}^{\rm (d)} + q E_P \vect B \sproduct \vect \BP \right) \tau_0 \ .
\end{multline}
The $2\times2$-matrices $\tau_i$ denote the Pauli matrices in electron-positron space, while $\sigma_i$ describes the usual spin degree of freedom. The magnetic dipole term is given according to (\ref{eq:OmegaRPp}) in terms of an interaction term between magnetic field $\vect B = \vect \nabla \xproduct \vect A$ and the fictitious momentum space magnetic field $\vect \BP$, to be specified below. We note that in this case, all the correction terms are equal for both electron and positron bands, albeit the term positron becomes only meaningful when all negative energy states are completely occupied.

The full matrix structure of the Berry connection, split into diagonal and off-diagonal parts, i.e. $\calA_{\vect P} = \calA_{\vect P}^{\rm (d)} \tau_0 + \calA_{\vect P}^{\rm (o)} \tau_y$, reads
\begin{align}
\calA_{\vect P}^{\rm (d)} &= \frac {c^2 \vect P \xproduct \sig }{ 2 E_P (E_P + m c^2) } \approx \frac{\lambda_c}{4\hbar} \frac {\vect P \xproduct \sig }{m c} \ , \\
\calA_{\vect P}^{\rm (o)} &= \frac {c \sig} {2E_P} - \frac{\vect P (\vect P \sig) \ c^3} {2 E_P^2 (E_P + mc^2)} \approx \frac {\lambda_c}{2\hbar} \sig  \ ,
\end{align}
where in the last step, we are dropping terms of order $ O(p/mc)^2$.

The kinetic variables for both positive and negative energy states read explicitly
\begin{align*}
\vect P &= \vect p - q\vect A(\vect r) - q \hbar \vect B \xproduct \Ad{\vect P} \ , \\
\vect R &= \vect r + \hbar \Ad{\vect P} \ ,
\end{align*}
so that the Berry curvature (\ref{eq:EffectiveFieldBP}) is also equal for both bands, and yields
\begin{align}
\vect \BP = -\frac{\hbar m c^4}{2E_P^3} \sig - \frac{\vect P (\vect P \sproduct \sig) \ \hbar c^4}{2E_P^3 (E_P + mc^2)} \approx -\frac{\lambda_c^2}{2\hbar} \sig \ ,
\end{align}
where again, in the last step we took the non-relativistic limit. Of course, the general expressions are still valid for arbitrary velocities, for example, in the opposite, the ultra-relativistic limit, we find
\begin{align*}
\vect \BP \stackrel{p \gg m c}{\rightarrow} - \frac {\vect P (\vect P \sproduct \sig) \hbar} {2 P^4} \ .
\end{align*}

Using the results (\ref{eq:FictitiousBRp}) - (\ref{eq:FictitiousThetaRPp}), we immediately find $\vect \ER = 0 = \vect \EP$ and
\begin{align}
\label{eq:FictitiousFieldsDirac}
\vect \BR &= q \vect B(\vect R) + q^2  \left(\vect \BP \sproduct \vect B \right) \vect B \nonumber\\
\ThetaRP_{ij} &= q \left[  \left( \vect \BP \sproduct \vect B \right) \delta_{ij} - \BP_i B_j \right] \ ,
\end{align}
where here, we used $\vect E(\vect R) = 0$, since we are already using the scalar potential $V(\vect R)$.

In terms of kinetic variables, the Hamiltonian $\bar\calH$ can now be rewritten (let us restrict ourselves to the positive energy branch, whose excitations correspond to electrons, so that $q = -e$), neglecting terms of order $\hdag^2$,
\begin{align}
\label{eq:DiracHamiltonianDiagonalized}
\bar\calH = E_P + V(\vect R) + q E_P \vect B(\vect R) \sproduct \vect \BP \ ,
\end{align}
so that in view of this, the spin-orbit interaction that appears in the rotated frame can be reinterpreted as being a result of the shift the kinetic position operator attains. However, we should keep in mind that this shift is gauge dependent and becomes physically meaningful only when integrated over the whole momentum space (which then is equivalent to the polarization) or in the form of the Berry curvature, viz the momentum space magnetic field appearing in the kinetic equations. In the non-relativistic limit, the magnetic dipole energy is just the usual Zeeman term (using (\ref{eq:HLinkOmega}) and (\ref{eq:DefinitionOmegaRP}))
\begin{align}
\label{eq:ZeemanTerm}
\bar\calH_\calM = - \vect \mu_s \sproduct \vect B \ ,
\end{align}
where we introduced the magnetic moment of the electron spin, $\vect \mu_s = -g_s \mu_{\rm B} \sig /2 = - \mu_{\rm B} \sig $, and we assume a g-factor of $2$ within the validity of the Dirac theory without quantum corrections from the radiative field, and $\mu_{\rm B} = \frac{e\hbar}{2m}$ is the Bohr magneton.

Using results (\ref{eq:SemiclassicalEquationsOfMotionR1}) and (\ref{eq:SemiclassicalEquationsOfMotionP1}) together with (\ref{eq:FictitiousFieldsDirac}) and (\ref{eq:DiracHamiltonianDiagonalized}), we obtain the operator equations of motion of a relativistic electron which moves in an electromagnetic field that is smooth on the  scale of the Compton wavelength $\lambda_c$. We do not explicitly write the expression here, however, our result is consistent with the work of Bliokh \cite{Bliokh2005}.

For the sake of our discussion, we will nevertheless specify the equations of motion in absence of a magnetic field,
\begin{align}
\label{eq:wavePacketDiracElectron}
\frac{\upd \vect R}{\upd t} &= \frac{\partial \bar \calH} {\partial \vect P} - \frac{\upd \vect P}{\upd t} \xproduct \vect \BP = \frac{\partial \bar \calH} {\partial \vect P} + \frac{\lambda_c^2}{2\hbar} \frac{\upd \vect P}{\upd t} \xproduct \sig \ , \\
\frac{\upd \vect P}{\upd t} &= -\frac{\partial \bar \calH} {\partial \vect R} = -e \vect E(\vect R) \ ,
\end{align}
whose anomalous velocity term is a factor $2$ larger than what one would obtain from the Hamiltonian (\ref{eq:DiracPauliHamiltonian}), naively taking the canonical variable $\vect r$ to be the physical position operator. Essentially, the same conclusion has been reached in Ref. \onlinecite{RevModPhysXiao2010} for the case of the relativistic Dirac equation.

The particle density is, according to results (\ref{eq:ParticleDensityDiagFrame}) and (\ref{eq:FictitiousFieldsDirac}), given by
\begin{align}
n(\vect R, t) = \int \frac{\upd^3 P}{(2\pi \hbar)^3} \  \Tr_4 \bar\rho \left(1 - 2q \vect B \sproduct \vect \BP \right) \ ,
\end{align}
and let us take the vacuum state with all negative energies filled, so that $\bar\rho = \Theta(-\tau_z E_P)$ and the first term gives the vacuum charge. However, from the correction factor proportional to the magnetic field one might conclude that a sufficiently strong magnetic field leads to a charge accumulation with respect to the vacuum state. This kind of charge accumulation is also present in topological insulators \cite{Nomura2010}. Here however, factors like $D = 1 - \Tr \ThetaRP = 1 - \frac {2\bar\calH_{\calM}} {m c^2}$ are relevant only for enormous magnetic fields of $\approx 10^{10} T$, so we can safely drop all corrections due to $\ThetaRP$.

It is also insightful to study the current density in the presence of the electromagnetic field, because it gives a direct meaning to the divergence-free current terms in (\ref{eq:CurrentDensityDiagFrame}). For the simplicity of our discussion, we restrict ourselves to the non-relativistic limit, where $\vect \BP =  -\frac{\lambda_c^2}{2\hbar} \sig = \frac {\vect \mu_s } {e mc^2}$ and $E_P = m c^2 + \vect P^2/2m$, and furthermore, we treat external fields $\vect E$ and $\vect B$ only to linear order. Since we want to study excitations of positive energy, we ignore the completely filled lower band.

Then, our charge density is $$n(\vect R, t) = -e \int \frac{\upd^\dm P}{(2\pi \hbar)^\dm} \  \bar\rho \ ,$$
so that the charge current density becomes with the help of (\ref{eq:CurrentDensityDiagFrame}) and (\ref{eq:SemiclassicalEquationsOfMotionR1}) as well as the definition of the spin magnetic moment (c.f. Eq. (\ref{eq:ZeemanTerm}))
\begin{multline}
\label{eq:DiracCurrentSemiclassical}
\vect j(\vect R, t) = -e\int \frac{\upd^\dm P}{(2\pi \hbar)^\dm} \ \frac{\vect P}{m} \bar\rho  - \frac{e}{mc^2}n(\vect R, t) \ \vect E \xproduct \vect \mu_s \\
- \vect \mu_s \xproduct \vect \nabla_{\scriptscriptstyle \!\!\vect R} \, n(\vect R, t) \ ,
\end{multline}
where the surface term vanishes since we assume our electron momentum to be peaked around some mean value $\vect P_c$. The density $\bar\rho$ can be explicitly determined from equation (\ref{eq:BoltzmannEquation}), here however, we only integrate this equation over all momenta and arrive at the continuity equation
\begin{align}
\partial_t n(\vect R, t) + \vect \nabla_{\scriptscriptstyle \!\!\vect R} \sproduct \vect j(\vect R, t) = 0 \ .
\end{align}
The first term in (\ref{eq:DiracCurrentSemiclassical}) is the usual definition of a non-relativistic current, and the second term is the anomalous velocity contribution due to the electric field. Like for the electron wave-packet considered in equation (\ref{eq:wavePacketDiracElectron}), using the non-relativistic Pauli Hamiltonian without the distinction between canonical and kinetic variables would yield an anomalous velocity term that is just half of the correct value. However, the last term seems unusual at first as it constitutes a persistent current which itself gives rise to a magnetic moment,
\begin{align}
\frac12 \int \upd^3 R \ \vect R \xproduct \vect j_{\rm per} &= - \frac12 \int \upd^3 R \ \vect R \xproduct (\vect \mu_s \xproduct \vect \nabla_{\scriptscriptstyle \!\!\vect R} \, n) \nonumber\\
&= \vect \mu_s \int \upd^3 R \ n(\vect R,t) = \calN \vect \mu_s \ ,
\end{align}
where $\calN=1$ is the number of electrons.
This result suggests that the notion of electron spin and this internal persistent current are just different interpretations of the same effect. For example, a magnetic field couples to this circular current via the magnetic moment it generates, and gives rise to the Zeeman energy (\ref{eq:ZeemanTerm}). This comes close to the original suggestion of an internal rotation of the electron by Uhlenbeck and Goudsmit \cite{Uhlenbeck1926}, yet unlike their idea, it is not the motion of a solid object, instead it is a genuine quantum phenomena, and thus does not suffer the same deficiencies concerning rotation velocities of the electron that would have to be faster than the speed of light.

As discussed before, the spin-orbit term in the Pauli Hamiltonian is {\it not} gauge invariant, it gives the correct energy spectrum, but not the correct equations of motion. That is why it has not been realized for a long time, as the Pauli Hamiltonian has been mainly put to test with electronic spectra in atoms or solid matter. To explicitly show this point, we go into a different rotated frame so that the Berry connection is changed to $\calA_{\vect p} \rightarrow \calA_{\vect p} + \hbar \vect \nabla_{\!\vect p} \chi(\vect p)$, and we get an additional term in the Pauli Hamiltonian $\hbar (\vect \nabla_{\!\vect p} \chi)  (\partial_{\vect r} V)$. Let us assume for the moment that we are working in momentum space and in the operator representation of quantum mechanics, where the Hamiltonian formally looks the same. Now, performing a local gauge transformation in momentum space by adding the phase factor $e^{i \chi(\vect p)}$ to the wave function, we obtain $e^{-i \chi(\vect p)}V(\vect r)e^{i \chi(\vect p)} = V(\vect r - \hbar \vect \nabla_{\!\vect p} \chi) = V(\vect r) - \hbar (\vect \nabla_{\!\vect p} \chi)  (\partial_{\vect r} V) + O(\hdag^2)$ which exactly cancels the additional term, and we are left with the original Hamiltonian. Usually, one enforces gauge-invariance in real space to obtain the electromagnetic field. This is a an example of gauge invariance in reciprocal space which leads to the phenomena of spin-orbit interaction, or the anomalous velocity.

In the above analysis, we needed to drop terms of order $\hbar^2$, though here, $\hbar$ is just a formal expansion parameter, the real relevant scale being the Compton wavelength $\lambda_c \equiv \frac{\hbar}{m c}$.
For example, here, we find that the order of $\hbar \calA^{\rm (d)}$ is $\frac {\hbar p}{4m^2 c^2} = \lambda_c \frac v c$ which is to be compared with our position coordinate, where one realizes that corrections are indeed small unless the potential $V(\vect r)$ varies strongly on the scale of the Compton wavelength $\lambda_c$. Consequently, the Hamiltonian (\ref{eq:DiracPauliHamiltonian}) still describes the correct quantum behavior for sufficiently smooth potentials and thus, the term semiclassical expansion seems misleading. However, for the case of the Coulomb delta-like potential of the nucleus of an atom it is obviously no longer true, and this leads to corrections like the Darwin term which is essential to understand atomic spectra \cite{SchwablQM2008}.

Sometimes, one interprets the terms appearing in the Pauli Hamiltonian as being a result of what one termed {\it Zitterbewegung}, and which one envisages as helical motion of the electron. Then the spin-orbit interaction and its physical consequences like the anomalous velocity can be understood as a result of this motion. Obviously, we do not observe such an oscillatory motion here in the effective theory, when the excitation is confined to a single band. Essentially, due to the Wigner transformation, this rapid oscillation has been transformed into energy space, and instead, the effects like anomalous velocity now appear in form of inter-band corrections.
We remark that in the case of massless Dirac-Fermions and if one considers excitations in the vicinity of the Dirac point, both positive and negative energy states have to be taken into account simultaneously so that excitations are not confined to positive or negative energies alone. In this scenario as for example in Graphene, one obtains those features termed {\it Zitterbewegung} \cite{Katsnelson2006}.

\section{Conclusions and Outlook}

In this work, we first studied a very general Hamiltonian that contains an additional matrix structure describing different bands, and the goal was to bring this Hamiltonian into band-diagonal form which has been achieved by performing a rotation in band-space. The diagonal representation is very practical when one wants to study the low-energy response of the system because then, only one or few degenerate bands are relevant. Since in the very general case this diagonalization involves the pair of non-commutating position $\vect r$ and momentum $\vect p$ operators, we performed the diagonalization perturbatively in $\hbar$ by using the Wigner representation. We investigated the Hamiltonian and physical observables in the rotated frame and how Berry curvatures emerge naturally during this diagonalization procedure. Essentially, Berry connections describe corrections to the canonical variables like position and momentum due to inter-band scattering.
This led to the distinction between canonical and kinetic variables, where the kinetic variables in the rotated frame are strongly linked to the canonical variables of the original theory. The canonical ones are, on the other side, responsible for the proper quantum structure due to their canonical commutation relations which defines the quantization. Therefore, both the canonical and kinetic variables are an important part of our effective description, and only then will it be consistent. We also established a link between gauge invariance in momentum space and the spin-orbit interaction, so that the Hamiltonian expressed in terms of canonical variables is enough to study energy spectrum of the system.

Having in mind a gauge-invariant description, we expressed the Hamiltonian and observables in terms of kinetic variables, which naturally leads to the appearance of Berry curvatures which are gauge invariant and describe various effects intrinsic to the band structure. In addition to Berry curvatures, we identified further gauge invariant objects which are related to circular, or persistent currents, which itself interact with magnetic fields or with each other and give for example rise to energy terms, like the Zeeman interaction.
For the Dirac equation, the electron is naturally delocalized in space and we found the appearance of an internal motion in the form of circular persistent currents giving rise to a magnetic dipole moment, and which exactly corresponds to the spin magnetic moment of the electron.

Using these kinetic variables, we formulated a Boltzmann transport equation that incorporates intrinsic effects expressed in terms of Berry curvatures. To this end, we consistently defined various quasi-probability densities which are connected by a conservation law: Liouville's theorem in phase space. These can be used to obtain physically meaningful densities and currents within the effective description. Furthermore, it is rather straightforward to include impurity scattering in the same manner as in Ref. \onlinecite{Wickles2009}, which is subject of future work. The results derived in this work provide a good starting point to transform the collision integral into the diagonalized frame, automatically incorporating the effect of an external electromagnetic field.

A related work by Wong and Tserkovnyak treated a general 2-band system by studying a quantum kinetic equation approach in the rotated frame \cite{Wong2011}. However, the rotation diagonalizes the Hamiltonian only to zeroth order in a gradient expansion, so that when studying gradient corrections the diagonalization is no longer exact. It is therefore no longer sufficient to consider only the diagonal part of the kinetic equation. To this end, it is necessary to take into account the off-diagonal elements which is in contrast to our formulation where already the rotation transformation includes gradient corrections.

Despite performing a formal expansion in $\hbar$, similar to treating the semiclassical limit, we never abandon the quantum description. One major advantage compared to other semiclassical approaches is that there is no need for a re-quantization (see Appendix \ref{sec.Requantization} for some details). In fact, our real expansion parameter might be a different one, like the Compton wave-length as we have seen in the case of the relativistic Dirac equation.
Also, our approach is systematic in the sense that we can go to arbitrary order in $\hbar$ or inter-band coupling. In this respect, it is also very interesting to look at terms second order in $\hbar$ which yield important contributions, for example the Darwin term in the case of the low-energy limit of the relativistic Dirac equation. Furthermore, new physical phenomena emerge at $O(\hbar^2)$, like the magneto-electric coupling in insulators which received new attention recently, also due to the discovery of topological insulators \cite{Essin2010, Coh2011}.

Then, one could study a variety of systems that involve both spin-orbit interaction (SOI) and an inhomogeneous and time-dependent magnetization. This is interesting for spin-orbit coupled semiconductors where one has different types like Rashba SOI, Dresselhaus SOI or in the case of strong SIO in III-V ferromagnetic semiconductors one can utilize the Luttinger Hamiltonian for hole transport. In these systems the electron or hole spin is intricately coupled to both momentum and orbital degrees of freedom. In addition, one can have non-magnetic and magnetic impurity scattering that adds additional complexity to the system and is naturally studied in terms of a collision integral within the Boltzmann approach. Thus, the formalism developed in this work seems practical to attach these kinds of systems in order to study various transport and dynamical properties. Finally, it would be interesting to study many particle effects like electron-electron interations.

This work was financially supported by the DFG though SFB 767 and SP1285.

\appendix

\section{Wigner Representation}
\label{ap:WignerRepresentation}
In the operator representation of quantum mechanics, we can write the canonical commutation relation (quantities with hats indicate operators in the original picture),
\begin{align}
\left[\hat {\vect \pi}_\mu, \hat {\vect x}^\nu \right] = i \hbar \delta_\mu^{\ \nu} \ .
\end{align}

In 4-component notation, the Wigner transformation explicitly reads
\begin{align}
\label{eq:WignerTransformFourComponent}
\int \upd^4 z \ e^{i \vect z \vect \pi / \hbar}  A\left(\vect x + \frac{\vect z}2, \vect x - \frac{\vect z}2\right) = A(\vect x, \vect \pi) \ ,
\end{align}
and the Moyal product obeys the axiom of associativity,
\begin{align}
A \Moyal B \Moyal C  = A \Moyal (B \Moyal C) = (A \Moyal B) \Moyal C \ ,
\end{align}
and also
\begin{align}
(A \Moyal B)^\dagger = B^\dagger \Moyal A^\dagger \ ,
\end{align}
however, the Moyal product is non-commutative, i.e. $A \Moyal B \neq B \Moyal A$. In fact, the latter property encodes the non-commutativity of operators in quantum theory which one can explicitly see by considering the commutator of two general observables dependent on $\vect x$ and $\vect \pi$ and which are band-diagonal,
\begin{align}
\left[A(\vect x, \vect \pi) \stackrel{\Moyal}{,}  B(\vect x, \vect \pi) \right] &= 2i A(\vect x, \vect \pi) \ \sin(\hbar \Lambda/2) \ B(\vect x, \vect \pi) \nonumber \\
&= i\hbar \left\{A, B \right\}_{\rm p} + O(\hbar^3) \ .
\end{align}
It becomes essentially an extended Poisson bracket in the semiclassical limit which reduces to the normal Poisson bracket when neither $A$ nor $B$ explicitly depend on $\epsilon$ which is usually the case.

The Wigner representation has the very practical feature of automatically symmetrizing operators, when one transforms back into the operator formulation, for example using the back transformation of (\ref{eq:WignerTransformFourComponent}),
\begin{align}
A(\vect x + \frac{\vect z}2, \vect x - \frac{\vect z}2) = \int \frac{\upd^4 \pi}{(2\pi \hbar)^4} \ e^{-i \vect z \vect \pi / \hbar} A(\vect x, \vect \pi) \ ,
\end{align}
one obtains
\begin{align}
x p \rightarrow \frac{1}{2} \left( \hat x \hat p + \hat p \hat x \right) \ ,
\end{align}
so what is usually required to be put in by hand in usual quantum mechanics in order to obtain Hermitian observables, comes out automatically.
Another, more direct way to see this is by using identities of the form
\begin{align}
\frac 12 \left( x \Moyal p + p \Moyal x \right) = x p \ .
\end{align}
The special operator ordering obtained in this manner is the so-called Wigner-Weyl ordering.

\section{Some Notes on Requantization}
\label{sec.Requantization}
In this section, we address the question of the quantum theoretical aspects of our semiclassical analysis. In particular, in literature there is the question of how to properly quantize the theory in the Lagrangian formalism, since one needs to identify the canonical pair of variables which is the starting point of the quantization. As mentioned previously, using wave packet analysis \cite{RevModPhysXiao2010}, one obtains the same set of equations of motion for the center of mass coordinates of the wave-packet as in (\ref{eq:SemiclassicalEquationsOfMotionR}) and  (\ref{eq:SemiclassicalEquationsOfMotionP}). Then one can find the Lagrangian, but one does not know anything about canonical variables, and in the general case, it is not always easy or possible to find the pair of canonical variables \cite{RevModPhysXiao2010}.

However, in our framework, we did not encounter such problems, since firstly, we are working in the Hamilton formalism, where one knows the canonical variables and secondly, our theory is quantized at any time, albeit neglecting terms of order $O(\hdag^2)$. This quantization is encoded in the structure of the Moyal product, or more explicitly in the special structure of the commutator in (\ref{eq:MoyalTransform}). Furthermore, without major effort, we can go back to the operator representation of quantum mechanics by undoing the Wigner transformation.

Since many of our expressions (for example (\ref{eq:TransformationRelationXi}), (\ref{eq:TransformationRelation}), (\ref{eq:CanonicToKineticPair}), etc.) are straightforwardly transformed back into the operator formulation of quantum mechanics, we formally obtain the same expression, except mixtures of $\vect r$ and $\vect p$ appear properly symmetrized (see Appendix \ref{ap:WignerRepresentation}). In fact, we retain the full quantum theory which gives correct results at least to order $\hbar$. One just has pay attention that the Wigner transformation is {\it only} with respect to the canonical pair of variables so kinetic variables have to be replaced accordingly.

%


\end{document}